%% file: main_ot.tex
\documentclass{article}\usepackage[paper=a4paper,top=20mm, bottom=30mm, inner=20mm, outer=20mm]{geometry}
\usepackage{natbib}

\input{Templates/StandardPackagesAndFormatting}
\input{Templates/StandardMathsMacros}
\graphicspath{{Figures/}}

\renewcommand\eqdef\equiv
\begin{document}
\input{Sections/FrontMatterArticle}

\input{Sections/Introduction}
\input{Sections/ProblemFormulation}
\input{Sections/CritWall}
\input{Sections/Collision}
\input{Sections/VolumeOutCritWall}
\input{Sections/Experiment}
\input{Sections/Conclusion}

\begin{myappendices}
	\input{Sections/Appendix_Const_Surface}
	\input{Sections/Appendix_BoundRiemann_CritWall}
	\input{Sections/Appendix_PerturbingShocks}
\end{myappendices}

\bibliographystyle{jfm}
\bibliography{Bibliography/Books,Bibliography/Mine,Bibliography/NumericalMethodsHyperbolic,Bibliography/ShallowWater,Bibliography/GravityCurrents,Bibliography/AnalysisHyperbolic,Local,Bibliography/ChannelHydraulics,Bibliography/StratifiedTurbulence}

\end{document}

%% file: Templates/StandardPackagesAndFormatting.tex
\usepackage{mathtools,amssymb,mathrsfs,bm,centernot,siunitx}
\usepackage[scr=boondox]{mathalpha}	
\usepackage{graphicx}
\usepackage{caption,subcaption,float}
\usepackage{tabularx,multirow}
\usepackage[inline]{enumitem}

\usepackage{xcolor}

\usepackage{calc,etoolbox}

\usepackage[inline]{enumitem}

\makeatletter

\@ifclassloaded{article}{
	\numberwithin{equation}{section}
	\numberwithin{figure}{section}
	\numberwithin{table}{section}
	
}{}

\@ifpackageloaded{hyperref}{}{\usepackage[hidelinks]{hyperref}}
\@ifpackageloaded{cleveref}{}{
	\usepackage[compress]{cleveref}

	\crefformat{section}{\S#2#1#3}
	\crefformat{subsection}{\S#2#1#3}
	\crefformat{subsubsection}{\S#2#1#3}
	\crefrangeformat{section}{\S\S#3#1#4 to~#5#2#6}
	\crefmultiformat{section}{\S#2#1#3}{ and~\S#2#1#3}{, \S#2#1#3}{ and~\S#2#1#3}
	
	\@ifclassloaded{jfm}{
		\crefname{figure}{figure}{figures}
		\Crefname{figure}{Figure}{Figures}
	}{
		\crefname{figure}{fig.}{figs.}
		\Crefname{figure}{Fig.}{Figs.}
	}
	
	\crefname{equation}{}{}
	\crefformat{equation}{(#2#1#3)}
	\crefmultiformat{equation}{(#2#1#3)}{ and (#2#1#3)}{, (#2#1#3)}{ and (#2#1#3)}
}

\@ifclassloaded{WileyNJD-v2}{}{
	\usepackage[title]{appendix}
	\newenvironment{myappendices}
	{
		\begin{appendices}
		\crefalias{section}{appsec}
	}{
		\end{appendices}
	}
	
	\crefname{appsec}{Appendix}{Appendices}
	\Crefname{appsec}{Appendix}{Appendices}
}

\usepackage{xspace}

\DeclareRobustCommand\onedot{\futurelet\@let@token\@onedot}
\def\@onedot{\ifx\@let@token.\else.\null\fi\xspace}

\def\eg{e.g\onedot} 
\def\ie{i.e\onedot} 
\def\cf{\textit{cf}\onedot}

\def\sf{s.f\onedot}

\@ifclassloaded{svjour3}{}{
	
}

\newcommand{\floattag}[1]{%
  \@namedef{the\@captype}{#1}%
  \@namedef{theH\@captype}{#1}%
  \addtocounter{\@captype}{-1}}

\makeatother

%% file: Sections/FrontMatterArticle.tex
\title{The unsteady overtopping of barriers by gravity currents and dam-break flows}

\author{Edward W.G. Skevington and Andrew J. Hogg}

\maketitle

\begin{abstract}
	The collision of a gravitationally-driven horizontal current with a  barrier following release from a confining lock is investigated using a shallow water model of the motion, together with a sophisticated boundary condition capturing the local interaction. The boundary condition permits several overtopping modes: supercritical, subcritical, and blocked flow. The model is analysed both mathematically and numerically to reveal aspects of the unsteady motion and to compute the proportion of the fluid trapped upstream of the barrier. Several problems are treated. Firstly, the idealised problem of a uniform incident current is analysed to classify the unsteady dynamical regimes. Then, the extreme regimes of a very close or distant barrier are tackled, showing the progression of the interaction through the overtopping modes. Next, the trapped volume of fluid at late times is investigated numerically, demonstrating regimes in which the volume is determined purely by volumetric considerations, and others where transient inertial effects are significant. For a Boussinesq gravity current, even when the volume of the confined region behind the barrier is equal to the fluid volume, $30\%$ of the fluid escapes the domain, and a confined volume $3$ times larger is required for the overtopped volume to be negligible. For a subaerial dam-break flow, the proportion that escapes is in excess of $60\%$ when the confined volume equals the fluid volume, and a barrier as tall as the initial release is required for negligible overtopping. Finally, we compare our predictions to experiments, showing a good agreement across a range of parameters.
\end{abstract}

%% file: Sections/Introduction.tex
\section{Introduction}

Many environmental fluid flows fall under the broad category of gravity currents, where a density difference between the current and its surrounding ambient drives predominantly horizontal fluid motion \citep{bk_Simpson_GCEL,bk_Ungarish_GCIv2}. Examples include flood events such as those caused by the collapse of a dam \citep{bk_Stoker_WW}, open channel hydraulics \citep{bk_Chow_OCH}, the spreading of toxic gas \citep{ar_Rottman_1985}, oil spillages \citep{ar_Hoult_1972}, cold fronts \citep{bk_Simpson_GCEL}, katabatic winds \citep{bk_Simpson_GCEL}, salinity currents \citep{ar_Simpson_1982}, and turbidity currents \citep{ar_Simpson_1982}. It is important to understand these fluid motions, especially for hazardous environmental currents, in order to ameliorate their effects on people, environment, and infrastructure.

Canonical problems in this area include `dam-break' and `lock-release' flows, where an initially quiescent and homogeneous layer of dense fluid is confined behind a barrier which is instantaneously removed to generate a current propagating along a horizontal bed. The two terms usually distinguish between regimes of different relative densities: for `dam-break' flows, the liquid current is substantially denser than the ambient air, so that the ambient can be neglected; whilst for `lock-release' flows the density of the current is comparable to that of the ambient, and often the two densities are assumed to be sufficiently close that the Boussinesq approximation is valid \citep{bk_Ungarish_GCIv2}. These flows are not only important for their applications, but are amenable to analytical study, and may be investigated using laboratory experiments \citep[\eg][]{bk_Simpson_GCEL,bk_Ungarish_GCIv2}.

In an environmental setting, gravity currents typically have a vastly greater horizontal length scale than vertical. This extreme aspect ratio means that, to leading order, the pressure is hydrostatic, and vertical integration of the governing equations yields the shallow water equations \citep[\eg][]{bk_Ungarish_GCIv2}. However, if the density of the current is comparable to that of the ambient, then the front of the gravity current forms a `head' at which it uplifts the ambient on a horizontal length scale of the same order as the depth of the current. Thus, hydrostatic pressure is not a valid approximation locally \citep{ar_Porcile_2022}, and the dynamics must be captured by alternative means. Many authors have pursued investigations of the local dynamics, for example \citet{ar_vonKarman_1940,ar_Benjamin_1968,ar_Simpson_1979,ar_Huppert_1980,ar_Borden_2013,ar_Konopliv_2016,ar_Ungarish_2017,ar_Ungarish_2018}. For the purposes of constructing a shallow water model, the crucial result is that $\bar{u} = \Fro \sqrt{g' \bar{h}}$, where $\bar{u}$ and $\bar{h}$ are the dimensional velocity and depth of the current local to the head, and $g'$ is the reduced gravity (see \cref{sec:overtop_PF}). The value of the Froude number $\Fro$ is determined either experimentally or from a simplified analytical model of the region local to the head. By selecting a value, or expression, for $\Fro$ we establish a dynamic boundary condition for the shallow water model of the current. When the ambient is deep in comparison to the current, and the density ratio between the two is constant, then the Froude number may be considered constant \citep{ar_Benjamin_1968}. In this regime it has been possible to establish a similarity solution for the two-dimensional motion that arises after a sufficient time following initiation \citep{ar_Fannelop_1972,ar_Hoult_1972,ar_Gratton_1994}, where the complete exact solution is also available \citep{ar_Hogg_2006}.

It is unusual for environmental gravity currents to flow across a horizontal bed, and often the dynamics of the current are at least partially controlled by the topography. A particularly dramatic interaction occurs when the topography transitions from horizontal to an adverse incline, the height of the crest being of order the depth of the current, while the streamwise extent of the incline and crest are much smaller than the length of the current (though potentially still larger than the depth of the current). Due to the extreme scales, the current interacts with this incline as a barrier, \ie the fluid will be abruptly slowed and deepen, and may pour over the barrier depending on its height and the local properties of the current.

\begin{figure}
	\centering
	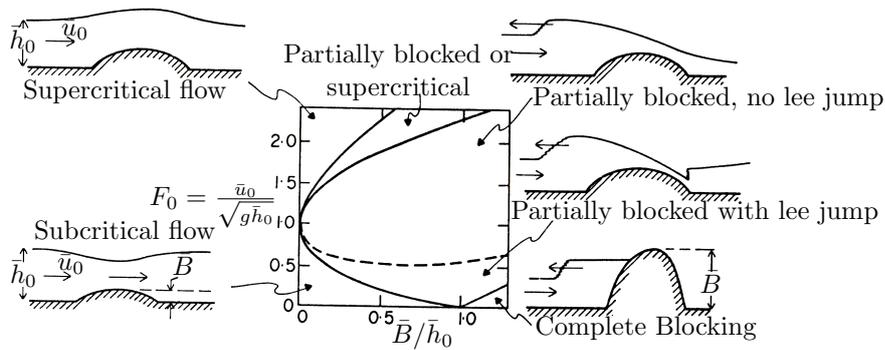
	\caption{Diagram from \citet[][with permission]{ch_Baines_1980} of the dynamical regimes in obstacle dragging experiments, modified for inclusion here. In the above figure $\bar{h}_0$ and $\bar{u}_0$ are the dimensional depth and velocity far upstream and downstream, and $\bar{B}$ is the dimensional height of the barrier.}
	\label{fig:ot:Baines_Davies_Regimes}
\end{figure}

Early work in this area focussed on the flow over a barrier, where the distal upstream and downstream flow conditions were the same. This investigation began with experimental work performed by \citet{ar_Long_1954}, where a barrier was placed at the bottom of the channel and instantaneously put in motion, so that in the frame of the obstacle the fluid impulsively mobilised. Analysis by \citet{ar_Long_1954} and \citet{ar_Houghton_1968} classified the dynamics, and this analysis was compiled by \citet{ch_Baines_1980} and \citet{bk_Baines_TESF} into \cref{fig:ot:Baines_Davies_Regimes}. This figure shows several dynamical regimes, wherein the fluid may simply flow over the obstacle subcritically or supercritically, or else shocks may be generated and the flow may become blocked. Interestingly, there is a parameter regime where multiple dynamical modes coexist, the history of the flow determining which is realised. It was shown numerically by \citet{ar_Pratt_1982} that it is possible to switch between the different dynamical regimes in a hysteresis loop by keeping the upstream and downstream flow conditions constant and altering the elevation of the obstacle with time. A similar hysteresis loop was found experimentally by \citet{ar_Lawrence_1987} keeping the obstacle height constant and varying the flow conditions.

The case of different distal upstream and downstream conditions has been investigated using the collision of a horizontally uniform current with a barrier. This problem was examined theoretically by \citet{ar_Rottman_1985}, who used a shallow water model to calculate the upstream depth for flow over a barrier with a vertical face, and \citet{ar_LaneSerff_1995} calculated the proportion of the incident flux that overtopped. Experimental verification of these results has been performed by \citet{ar_Pari_2017}. Additionally, \citet{ar_Ermanyuk_2005} experimentally measured the forces exerted on the barrier, and further investigation of this scenario by \citet{ar_GonzalezJuez_2009a} modelled the motion using DNS (direct numerical simulations). Both DNS and `box model' analysis (horizontally uniform shallow water) were used in \citet{ar_GonzalezJuez_2009b} to determine the flow downstream of the barrier. The classification of dynamical regimes for collision is similar to, but distinct from, that in \cref{fig:ot:Baines_Davies_Regimes}; see \citet{ar_Cozzolino_2014} and our \cref{sec:overtop_close}.

We now turn our attention to unsteady incident currents, which have been less extensively studied. \citet{ar_Greenspan_1978} investigated the collision of a `dam-break' flow with a barrier. Their study established an asymptotic expression for the initial deepening of the fluid layer adjacent to the barrier following the collision. This analysis has been significantly extended by \citet{ar_Skevington_F002_SemiInf_Wall} to provide a quasi-analytical calculation of the reflection of the flow by the barrier, revealing the rate at which the fluid deepens and the energy losses by the reflected bore. \citet{ar_Greenspan_1978} also provided numerical simulations of their model of shallow water overtopping along with comparisons to experiment, though their model assumes that the surface elevation of the fluid is constant on the incline of the barrier, and this requires an unphysical assumption about the dissipation of energy as the fluid flows up the incline (see \cref{sec:overtop_app_constsurface}).

In the present work the focus is on the unsteady collision of currents from lock-release and dam-break initial conditions with a barrier. In particular, we explore how the overtopping dynamics change over time, that is the transitions between subcritical, supercritical, and blocked flow, and how the evolution of these dynamics depends on the dimensionless parameters of the problem. In addition, we are interested in how the barrier traps fluid, and we calculate the portion of a finite release that remains upstream of the barrier.

The paper is structured as follows. In \cref{sec:overtop_PF} we develop a shallow water model of the collision, wherein the barrier is represented by a boundary condition that captures the energy conserving interaction, modifying the condition by \citet{ar_Cozzolino_2014} and generalising the condition by \citet{ar_Skevington_F001_Draining} (see \cref{sec:overtop_app_constsurface} for a discussion of conditions used by other authors). This boundary condition is studied in detail in \cref{sec:overtop_close} (with additional details in \cref{sec:overtop_app_Riemann}), classifying the type of solution seen locally to the boundary. This classification is similar to that in \cref{fig:ot:Baines_Davies_Regimes}, except we impose supercritical flow beyond the barrier, making the solution unique. We then proceed to classify the time evolving states seen in our dynamic collision process, when the barrier is sufficiently close to the release so the current is unaffected by its finite extent (\cref{sec:overtop_collision_close}), and when the barrier is sufficiently far so the current is in similarity form (\cref{sec:overtop_collision_far} using results from \cref{sec:overtop_PerturbShock}). We present numerical simulations of the shallow water model in \cref{sec:overtop_simulation}, where we examine how the portion of fluid that overtops the barrier depends on the parameters of the problem. Finally, in \cref{sec:overtop_experiment} we compare our predictions with the experimental results of \citet{ar_Greenspan_1978} and \citet{ar_GonzalezJuez_2009b}, showing a good correspondence and validating the model. We conclude in \cref{sec:conclusion}.

%% file: Figures/Baines_Davies_Regimes.eps_tex
\begingroup%
  \makeatletter%
  \providecommand\color[2][]{%
    \errmessage{(Inkscape) Color is used for the text in Inkscape, but the package 'color.sty' is not loaded}%
    \renewcommand\color[2][]{}%
  }%
  \providecommand\transparent[1]{%
    \errmessage{(Inkscape) Transparency is used (non-zero) for the text in Inkscape, but the package 'transparent.sty' is not loaded}%
    \renewcommand\transparent[1]{}%
  }%
  \providecommand\rotatebox[2]{#2}%
  \newcommand*\fsize{\dimexpr\f@size pt\relax}%
  \newcommand*\lineheight[1]{\fontsize{\fsize}{#1\fsize}\selectfont}%
  \ifx\svgwidth\undefined%
    \setlength{\unitlength}{294.30851073bp}%
    \ifx\svgscale\undefined%
      \relax%
    \else%
      \setlength{\unitlength}{\unitlength * \real{\svgscale}}%
    \fi%
  \else%
    \setlength{\unitlength}{\svgwidth}%
  \fi%
  \global\let\svgwidth\undefined%
  \global\let\svgscale\undefined%
  \makeatother%
  \begin{picture}(1,0.43341953)%
    \lineheight{1}%
    \setlength\tabcolsep{0pt}%
    \put(0,0){\includegraphics[width=\unitlength]{Baines_Davies_Regimes.eps}}%
    \put(0.1424916,0.31574195){\color[rgb]{0,0,0}\makebox(0,0)[t]{\lineheight{1.25}\smash{\begin{tabular}[t]{c}Supercritical flow\end{tabular}}}}%
    \put(0.1424916,0.13606732){\color[rgb]{0,0,0}\makebox(0,0)[t]{\lineheight{1.25}\smash{\begin{tabular}[t]{c}Subcritical flow\end{tabular}}}}%
    \put(0.01352995,0.37946252){\color[rgb]{0,0,0}\makebox(0,0)[t]{\lineheight{1.25}\smash{\begin{tabular}[t]{c}$\bar{h}_0$\end{tabular}}}}%
    \put(0.07933225,0.39517685){\color[rgb]{0,0,0}\makebox(0,0)[t]{\lineheight{1.25}\smash{\begin{tabular}[t]{c}$\bar{u}_0$\end{tabular}}}}%
    \put(0.07433296,0.09641652){\color[rgb]{0,0,0}\makebox(0,0)[t]{\lineheight{1.25}\smash{\begin{tabular}[t]{c}$\bar{u}_0$\end{tabular}}}}%
    \put(0.21613642,0.0892365){\color[rgb]{0,0,0}\makebox(0,0)[t]{\lineheight{1.25}\smash{\begin{tabular}[t]{c}$\bar{B}$\end{tabular}}}}%
    \put(0.88478391,0.06670551){\color[rgb]{0,0,0}\makebox(0,0)[t]{\lineheight{1.25}\smash{\begin{tabular}[t]{c}$\bar{B}$\end{tabular}}}}%
    \put(0.33528958,0.17850093){\color[rgb]{0,0,0}\makebox(0,0)[rt]{\lineheight{1.25}\smash{\begin{tabular}[t]{r}$F_0=\frac{\bar{u}_0}{\sqrt{g \bar{h}_0}}$\end{tabular}}}}%
    \put(0.51960849,0.00366684){\color[rgb]{0,0,0}\makebox(0,0)[t]{\lineheight{1.25}\smash{\begin{tabular}[t]{c}$\bar{B}/\bar{h}_0$\end{tabular}}}}%
    \put(0.01423508,0.08002585){\color[rgb]{0,0,0}\makebox(0,0)[t]{\lineheight{1.25}\smash{\begin{tabular}[t]{c}$\bar{h}_0$\end{tabular}}}}%
    \put(0.66241651,0.01316578){\color[rgb]{0,0,0}\makebox(0,0)[lt]{\lineheight{1.25}\smash{\begin{tabular}[t]{l}Complete Blocking\end{tabular}}}}%
    \put(0.6319786,0.16103539){\color[rgb]{0,0,0}\makebox(0,0)[lt]{\lineheight{1.25}\smash{\begin{tabular}[t]{l}Partially blocked with lee jump\end{tabular}}}}%
    \put(0.65950242,0.30668239){\color[rgb]{0,0,0}\makebox(0,0)[lt]{\lineheight{1.25}\smash{\begin{tabular}[t]{l}Partially blocked, no lee jump\end{tabular}}}}%
    \put(0.48943993,0.36151672){\color[rgb]{0,0,0}\makebox(0,0)[t]{\lineheight{1.25}\smash{\begin{tabular}[t]{c}Partially blocked or \\supercritical\end{tabular}}}}%
  \end{picture}%
\endgroup%

%% file: Sections/ProblemFormulation.tex
\section{Problem formulation}	\label{sec:overtop_PF}

\subsection{Governing equations}

We model the two dimensional motion of relatively dense fluid following instantaneous release from a quiescent state in a lock of dimensional depth $Z$ and length $X$ surrounded by a relatively deep and dynamically passive ambient. This two-dimensional model is equivalent to a laterally uniform three-dimensional flow along a rectangular channel. Driven by its density difference with the ambient, the released fluid flows along the underlying impermeable boundary forming a gravity current. On the assumption that the motion is shallow ($Z \ll X$) and predominantly parallel with the basal boundary, the pressure adopts a hydrostatic distribution to leading order. Neglecting mixing with the ambient and assuming the inclination of the bed is small, the dimensionless governing equations are the non-linear shallow water equations \citep[\eg][]{ar_Peregrine_1972,bk_Ungarish_GCIv2}
\sidebysidesubequations{eqn:ot:SW_sys}
{\pdv{h}{t} + \pdv{}{x} (uh) = 0,}{eqn:ot:SW_sys_mass}
{and}
{\pdv{}{t}(uh) + \pdv{}{x} \ppar*{ u^2 h + \frac{h^2}{2} } = -h \dv{b}{x},}{eqn:ot:SW_sys_mom}
which represent conservation of mass and the balance of momentum respectively. We have non-dimensionalised the horizontal coordinate $x$ by  $X$; the time $t$ by $\frac*{X}{\sqrt{g'Z}}$; the depth of the fluid $h(x,t)$ and bed elevation $b(x)$ by $Z$; and the velocity of the fluid $u(x,t)$ by $\sqrt{g'Z}$, where $g' \eqdef \frac*{g (\rho_f - \rho_a)}{\rho_f}>0$ is the reduced gravity, $g$ is the acceleration due to gravity, $\rho_f$ is the density of the flowing layer, and $\rho_a$ is the density of the dynamically passive ambient. When considering volumes of fluid, we scale by the width of the implied rectangular channel, and therefore calculate the area in the two-dimensional model.

In regions where $u$ and $h$ are continuous \cref{eqn:ot:SW_sys} may be written in terms of the characteristic invariants $\alpha \eqdef u + 2 h^{1/2}$ and $\beta \eqdef u - 2 h^{1/2}$ \citep[\eg][]{bk_Stoker_WW}
\begin{subequations}\label{eqn:ot:SW_characteristics_bed}\begin{align}
	\pdv{\alpha}{t} + \lambda \pdv{\alpha}{x} + \dv{b}{x} &= 0,
	&&\text{where}&
	\lambda &\eqdef u + h^{1/2},
	\label{eqn:ot:SW_characteristics_bed_alpha}
	\\
	\pdv{\beta}{t} + \mu \pdv{\beta}{x} + \dv{b}{x} &= 0,
	&&\text{where}&
	\mu &\eqdef u - h^{1/2}.
	\label{eqn:ot:SW_characteristics_bed_beta}
\end{align}\end{subequations}
We term the curves $\dv*{x}{t} = \lambda$ as the $\alpha$-characteristics, and $\dv*{x}{t} = \mu$ as the $\beta$-characteristics. When $b$ is constant then $\alpha$ is constant on $\alpha$-characteristics and $\beta$ is constant on $\beta$-characteristics. The flow is subcritical when the characteristics move in opposite directions, that is $\lambda \mu < 0$ or equivalently $\abs{u}<h^{1/2}$; supercritical when the characteristics move in the same direction, that is $\lambda \mu > 0$ or equivalently $\abs{u}>h^{1/2}$; and critical when one of the characteristics is stationary, that is $\lambda \mu = 0$ or equivalently $\abs{u}=h^{1/2}$.

Along shock curves $x = x_s(t)$ where the solution is discontinuous (but $b$ is continuous) we enforce conservation of the mass and momentum fluxes \citep[\eg][]{bk_Stoker_WW}, given by
\sidebysidesubequations{eqn:ot:SW_shock_con_general}
{\shock*{(u-s) h} = 0,}{eqn:ot:SW_shock_con_general_mass}
{and}
{\shock*{(u-s)^2 h + \frac{h^2}{2}} = 0}{eqn:ot:SW_shock_con_general_momentum}
respectively, where $s \eqdef \dv*{x_s}{t}$ and $\shock{f(x,t)} \eqdef f(x_s^+(t),t) - f(x_s^-(t),t)$. In addition, we require that any shocks absorb characteristics of one family, consistent with the Lax Entropy condition \citep{ar_Lax_1957}. Shocks for which $\lambda^- > s > \lambda^+$ are termed $\alpha$-shocks and satisfy $h^->h^+$, $s>u^->u^+$, whilst those with $\mu^- > s > \mu^+$ are termed $\beta$-shocks and satisfy $h^-<h^+$, $u^->u^+>s$.

\begin{figure}
	\centering
	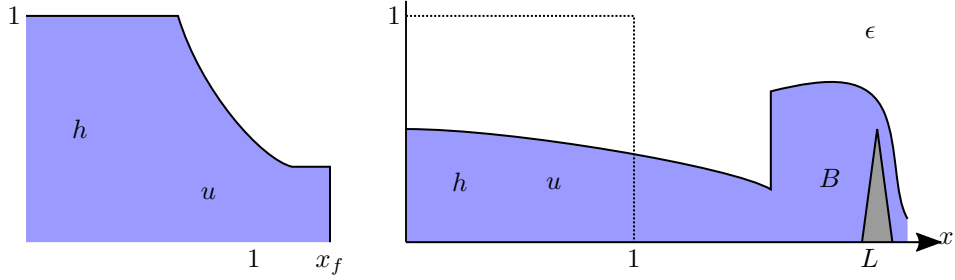
	\caption{The flow configuration, showing the dimensionless variables. The left figure shows the early time release with a moving front, and the right the flow over the barrier during subcritical overtopping. The vertical scale is exaggerated and the horizontal axis non-linearly scaled to show the current and the barrier. In dotted is the initial release, in solid is the depth field, and dashed lines are used to indicate measurements for the barrier.}
	\label{fig:ot:Configuration}
\end{figure}

The domain is $0 \leq x \leq x_f(t)$, and the initial conditions are $h(x,0)=1$, $u(x,0)=0$, $x_f(0)=1$. The bed is horizontal up until the barrier at $x=L_1 \geq 1$, that is $b(x) = 0$ for $0 \leq x < L_1$, and the fluid is confined to the rear by an insurmountable barrier imposing $u(0,t)=0$, which we call the back-wall. The motion during $t>0$ is due to the removal of the lock at $x=1$ which produces a moving front obeying \citep{ar_Benjamin_1968}
\begin{align} \label{eqn:front_condition}
	\dot{x}_f \eqdef \dv{x_f}{t} = u &= \Fro \, h^{1/2}
	&&\text{at}&
	x &= x_f (t)
\end{align}
where $\Fro$ is the frontal Froude number, which is dependent on the density ratio $\rho_f/\rho_a$.
By this definition of $\Fro$, Boussinesq currents ($\rho_f/\rho_a \approx 1$) take a theoretically derived value of $\sqrt{2}$ \citep{ar_Benjamin_1968,ar_Ungarish_2018} and experimentally measured value of $1.19$ \citep{ar_Huppert_1980}, while dam-break flows operate in the limit $\Fro \to \infty$ and the front corresponds to vanishing height $h(x_f,t)=0$. In our analysis we will assume that $\Fro>0$ is constant. The configuration is shown in \cref{fig:ot:Configuration}.

A barrier is located between $L_1 < x <L_2$, and over this interval $\dv*{b}{x} \geq 0$ with the crest at elevation $B \eqdef b(L_2)$, thus the overall angle of the incline is $\theta \eqdef \arctan \ppar*{ \frac*{B Z}{\epsilon X} }$ where $\epsilon \eqdef L_2 - L_1$. We assume the barrier is steep \emph{relative} to the scales of the current, so that $1 \ll B/\epsilon = (X/Z) \tan \theta$. Note that this does not prevent $\theta$ from being small because $Z \ll X$. For the current to be able to surmount the barrier we require that $B \lesssim 1$ (it will turn out that we require $B<2$, \cref{sec:overtop_collision_close}), so  $0 < \epsilon \ll 1$.

Because the barrier is short relative to the length of the current, we neglect the $\order{\epsilon}$ effects of its finite extent by including it as a boundary condition at $x=L$. That is, in the model we analyse in all subsequent sections, the bed elevation is set to zero across all of $0<x<L$, where $x=L$ is the location of the boundary condition which models the barrier. In principle we may set the boundary to be at any location $L_1 < L < L_2$, but to minimise the error in measuring the volume of fluid that escapes the domain (\cref{sec:overtop_simulation}) we ensure that the confined volume (that is, the volume of the region behind the barrier) $V_c$ is the same as in the configuration with $\epsilon$ non-vanishing, where
\begin{align} \label{eqn:confined_vol}
	V_c \eqdef \int_0^{L_2} B - b(x) \dd{x}.
\end{align}
That is, we choose $L \eqdef V_c/B$.

In what follows we denote the velocity and depth of the fluid at the base of the barrier by $u_b$ and $h_b$. When considering the configuration with non-vanishing $\epsilon$ these are expressed by $u_b(t) = u(L_1,t)$ and $h_b(t) = h(L_1,t)$, whereas at the crest of the barrier the velocity and depth are $u_c(t) = u(L_2,t)$ and $h_c(t) = h(L_2,t)$. The overtopping model developed below provides a dynamical connection between the states at the base and the crest and leads to conditions on $u_b$ and $h_b$. For our simplified model of the interaction between the current and barrier, which is accurate in the limit $\epsilon \rightarrow 0^+$, we impose these as conditions at $x=L$, that is the consideration of the barrier provides constraints on $u_b(t) = u(L,t)$ and $h_b(t) = h(L,t)$. We will also utilise the flow-state just upstream of the barrier, in the simplified model this is given by $u_l(t) = \lim_{x \to L^-} u(x,t)$ and $h_l(t) = \lim_{x \to L^-} h(x,t)$.

To develop a boundary condition that captures the effect of the barrier on the fluid, we use its relative shortness, $\epsilon \ll 1$, which implies the time scale of the fluid flow local to the barrier is vastly smaller than that of the flow in the bulk. Thus, the dynamics are quasi-static and steady state analysis may be applied. Following \citet{ar_Long_1954,ar_Long_1970} and \citet{bk_Baines_TESF}, we assume the flow to be continuous between the base and the crest (shocks are known to be unstable, see \citealt{ar_Baines_2003}). Thus, the dimensionless energy density $E \eqdef \textfrac{1}{2} u^2 + h + b$ and volume flux $q \eqdef uh$ are constant in space and time, with $q_b$, $E_b$ the values evaluated at the base of the barrier, and $q_c$, $E_c$ at the crest. At constant $q$ and $b$ the energy takes its minimal value at critical flow, $h = q^{2/3}$, at which $E = \textfrac{3}{2} q^{\frac*{2}{3}} + b$. For larger energies there are two possible corresponding values of $u$ and $h$, a subcritical state and a supercritical state. For the fluid to overtop the barrier the energy at the crest $E_c$ must be at least the critical energy there,
\begin{subequations}
\begin{flalign}
&&&
	E_c \geq \eval*{E_c}_{\text{crit}} = \textfrac{3}{2} q_c^{\frac*{2}{3}} + B = \textfrac{3}{2} q_b^{\frac*{2}{3}} + B,	&\\
&\text{thus}&&
	\label{eqn:ot:BC_energy_loss}
	\Delta E_b \eqdef E_b - \eval*{E_c}_{\text{crit}} = \Delta E (u_b,h_b,B) \geq 0	&\\
&\text{where}&& \label{eqn:ot:energy_discrep}
	\Delta E (\mathscr{U},\mathscr{H},\mathscr{B}) \eqdef \textfrac{1}{2}\mathscr{U}^2 + \mathscr{H} - \textfrac{3}{2} (\mathscr{U} \mathscr{H})^{\frac*{2}{3}} - \mathscr{B}	&
\end{flalign}
\end{subequations}
is the energy discrepancy. Note that this analysis assumes there is no energy dissipation as the fluid flows over the obstacle, which is justifiable for shallow sloped barriers. For steeper slopes, \cite{ar_Skevington_F001_Draining} showed, for open channel flows, how to account for the small amount of dissipation present. This is not pursued here.

We assume that the flow downstream of the barrier is downhill and supercritical, and conclude that there are three possible modes that can be exhibited:
\begin{description}
	\item[Supercritical overtopping ] The flow on the incline is supercritical, $u_b > h_b^{1/2}$, and satisfies $\Delta E_b  > 0$, which imposes no boundary condition on the bulk flow.
	\item[Subcritical overtopping ] The flow on the incline is subcritical and transitions to supercritical  beyond the crest. Thus, it is critical at the crest, and the appropriate boundary condition is $\Delta E_b = 0$, taking the solution in $0 \leq u_b \leq h_b^{1/2}$.
	\item[Blocked flow ] If $\Delta E_b < 0$ then the fluid cannot overtop the barrier, and the appropriate boundary condition is $q_b = 0$.
\end{description}

The strict inequality on energy for the case of supercritical overtopping ensures the stability of the continuous flow on the incline. For supercritical flow with $\Delta E_b  = 0$, a small perturbation to the values of $u_b$, $h_b$ may cause $\Delta E_b$ to become negative, at which point an upstream propagating bore will be generated, transitioning to one of the other two modes.

Identifying which of the three modes is selected depends on the flow local to the barrier $u_l$, $h_l$. As discussed by \citet{ar_Cozzolino_2014}, there is a choice of whether to permit supercritical overtopping in some cases. By the experimental measurements of \citet{ar_Greenspan_1978} we expect that slopes of $\theta = 60\degree$, perhaps greater, are able to produce a supercritical jet of overtopping fluid for an incident dam-break flow, and to capture this flow regime we permit supercritical overtopping when $u_l > h_l^{1/2}$ and $\Delta E_l \eqdef \Delta E (u_l,h_l,B) > 0$. Otherwise, if a solution exists to imposing subcritical overtopping then this boundary condition is used, else we impose blocked flow. Collectively, we term all three modes and the method of selection the \emph{critical barrier boundary condition}, and we examine the consequence of its imposition in \cref{sec:overtop_close}.

Works by other authors \citep[\eg][]{ar_Greenspan_1978,ar_Rottman_1985} have used a different outflow condition where a constant surface elevation is imposed rather than conservation of energy. In \cref{sec:overtop_app_constsurface} we show that the surface elevation model is unphysical, and that the conservation of energy condition is the physically permitted condition that minimises the change in surface elevation.

The model is validated by comparison to experiment in \cref{sec:overtop_experiment}, showing a good correspondence. However, we first analyse the properties and predictions of the model in \cref{sec:overtop_close,sec:overtop_collision_extreme,sec:overtop_simulation} to give context to the comparison.

\subsection{Numerical methods}

We simulate using the transformed shallow water system presented in \citet{ar_Skevington_N001_Boundary}. The numerical method employs the central upwind scheme by \citet{ar_Kurganov_2002}. This is a finite volume scheme which requires reconstruction in each cell, for this purpose we use the suppressed minmod limiter from \citet{ar_Skevington_N003_Recon_Test}. Boundary conditions are implemented using techniques developed in \citet{ar_Skevington_N001_Boundary}, which ensure the solution is locally continuous at the boundaries for almost all time, and that there is no drift off error in the algebraic boundary conditions.

%% file: Figures/Configuration_Wall.eps_tex
\begingroup%
  \makeatletter%
  \providecommand\color[2][]{%
    \errmessage{(Inkscape) Color is used for the text in Inkscape, but the package 'color.sty' is not loaded}%
    \renewcommand\color[2][]{}%
  }%
  \providecommand\transparent[1]{%
    \errmessage{(Inkscape) Transparency is used (non-zero) for the text in Inkscape, but the package 'transparent.sty' is not loaded}%
    \renewcommand\transparent[1]{}%
  }%
  \providecommand\rotatebox[2]{#2}%
  \newcommand*\fsize{\dimexpr\f@size pt\relax}%
  \newcommand*\lineheight[1]{\fontsize{\fsize}{#1\fsize}\selectfont}%
  \ifx\svgwidth\undefined%
    \setlength{\unitlength}{368.50393701bp}%
    \ifx\svgscale\undefined%
      \relax%
    \else%
      \setlength{\unitlength}{\unitlength * \real{\svgscale}}%
    \fi%
  \else%
    \setlength{\unitlength}{\svgwidth}%
  \fi%
  \global\let\svgwidth\undefined%
  \global\let\svgscale\undefined%
  \makeatother%
  \begin{picture}(1,0.26923077)%
    \lineheight{1}%
    \setlength\tabcolsep{0pt}%
    \put(0,0){\includegraphics[width=\unitlength]{Configuration_Wall.eps}}%
    \put(0.96219113,0.02119317){\color[rgb]{0,0,0}\makebox(0,0)[lt]{\lineheight{1.25}\smash{\begin{tabular}[t]{l}$x$\end{tabular}}}}%
    \put(0.57261124,0.07686019){\color[rgb]{0,0,0}\makebox(0,0)[t]{\lineheight{1.25}\smash{\begin{tabular}[t]{c}$u$\end{tabular}}}}%
    \put(0.86304216,0.07904082){\color[rgb]{0,0,0}\makebox(0,0)[rt]{\lineheight{1.25}\smash{\begin{tabular}[t]{r}$B$\end{tabular}}}}%
    \put(0.46988343,0.07503932){\color[rgb]{0,0,0}\makebox(0,0)[lt]{\lineheight{1.25}\smash{\begin{tabular}[t]{l}$h$\end{tabular}}}}%
    \put(0.41784721,0.24542697){\color[rgb]{0,0,0}\makebox(0,0)[rt]{\lineheight{1.25}\smash{\begin{tabular}[t]{r}$1$\end{tabular}}}}%
    \put(0.65380971,-0.00188376){\color[rgb]{0,0,0}\makebox(0,0)[t]{\lineheight{1.25}\smash{\begin{tabular}[t]{c}$1$\end{tabular}}}}%
    \put(0.89227125,-0.00188376){\color[rgb]{0,0,0}\makebox(0,0)[t]{\lineheight{1.25}\smash{\begin{tabular}[t]{c}$L$\end{tabular}}}}%
    \put(0.89282168,0.23128322){\color[rgb]{0,0,0}\makebox(0,0)[t]{\lineheight{1.25}\smash{\begin{tabular}[t]{c}$\epsilon$\end{tabular}}}}%
    \put(0.22304048,0.06734701){\color[rgb]{0,0,0}\makebox(0,0)[t]{\lineheight{1.25}\smash{\begin{tabular}[t]{c}$u$\end{tabular}}}}%
    \put(0.08526804,0.1288855){\color[rgb]{0,0,0}\makebox(0,0)[lt]{\lineheight{1.25}\smash{\begin{tabular}[t]{l}$h$\end{tabular}}}}%
    \put(0.03323182,0.24542701){\color[rgb]{0,0,0}\makebox(0,0)[rt]{\lineheight{1.25}\smash{\begin{tabular}[t]{r}$1$\end{tabular}}}}%
    \put(0.26919432,-0.00188376){\color[rgb]{0,0,0}\makebox(0,0)[t]{\lineheight{1.25}\smash{\begin{tabular}[t]{c}$1$\end{tabular}}}}%
    \put(0.3461174,-0.00188376){\color[rgb]{0,0,0}\makebox(0,0)[t]{\lineheight{1.25}\smash{\begin{tabular}[t]{c}$x_f$\end{tabular}}}}%
  \end{picture}%
\endgroup%

%% file: Sections/CritWall.tex
\section{Boundary-Riemann problem adjacent to a barrier}	\label{sec:overtop_close}

\begin{figure}
	\centering
	\includegraphics{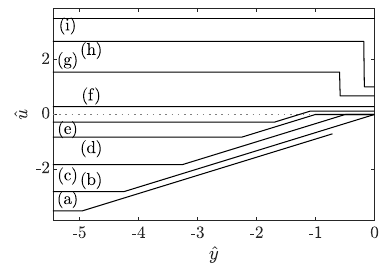}
	\includegraphics{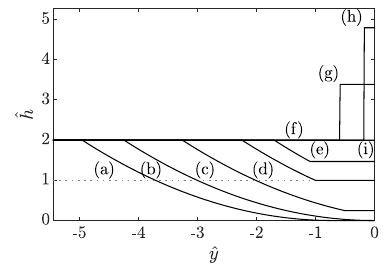}
	\caption{The velocity, $\hat{u}$, and depth, $\hat{h}$, as functions of $\hat{x}/\hat{t}$ for those initial uniform states $(\hat{u}_i,\hat{h}_i)$ marked on \cref{fig:ot:CritWall_BndRie} with $\hat{h}_i=2$. In dotted lines are plotted $\hat{h} = 1$, the height of the barrier, and $\hat{u} = 0$. The plotted solutions are at the following initial Froude numbers: (a) $F_i = -2.5$, (b) $F_i = -2$, (c) $F_i = -1.3$, (d) $F_i = -0.5858$, (e) $F_i = -0.2$, (f) $F_i = 0.2047$, (g) $F_i = 1.1$, (h) $F_i = 1.8963$, (i) $F_i = 2.5$.}
	\label{fig:ot:CritWall_BndRie_Prof1}
	\includegraphics{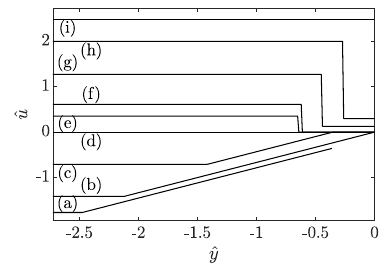}
	\includegraphics{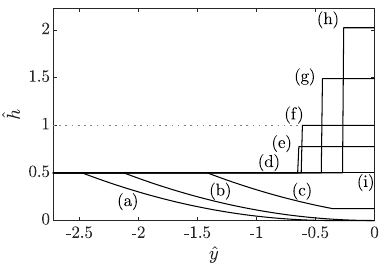}
	\caption{The velocity, $\hat{u}$, and depth, $\hat{h}$, as functions of $\hat{x}/\hat{t}$ for those initial uniform states $(\hat{u}_i,\hat{h}_i)$ marked on \cref{fig:ot:CritWall_BndRie} with $\hat{h}_i=0.5$. In dotted lines are plotted $\hat{h} = 1$, the height of the barrier, and $\hat{u} = 0$. The plotted solutions are at the following initial Froude numbers: (a) $F_i = -2.5$, (b) $F_i = -2$, (c) $F_i = -1$, (d) $F_i = 0$, (e) $F_i = 0.5$, (f) $F_i = 0.8660$, (g) $F_i = 1.8$, (h) $F_i = 2.8284$, \\(i) $F_i = 3.5$.}
	\label{fig:ot:CritWall_BndRie_Prof2}
\end{figure}

In this section we analyse the flow predicted by the governing equations \cref{eqn:ot:SW_sys} and the critical barrier condition at the right end of the domain when a uniform initial state is imposed. Mathematically, this configuration corresponds to a boundary-Riemann problem and its solution provides considerable insight into the manner in which the boundary condition controls the flow. This is the same problem as considered by \citet{ar_Cozzolino_2014}, though we choose supercritical outflow whenever multiple solutions exists. We present the solution to the boundary-Riemann problem in this section because it is used extensively later. This initial configuration may alternatively be viewed as a transformation of the problem posed in \cref{sec:overtop_PF} by introducing coordinates $\hat{x} = {(x-L)}/{\hat{\delta}}$, $\hat{t} = {(t-t_0) B^{1/2}}/{\hat{\delta}}$ so that the local behaviour at time $t_0$ can be examined over a region of size $\hat{\delta}$ where $\epsilon \ll \hat{\delta} \ll 1$ so that the variables are initially constant and the domain is $\hat{x} \leq 0$, $\hat{t} \geq 0$. It is convenient to rescale the dependent variables as
\begin{align}\label{eqn:ot:overtop_close_ic}
	\hat{h} &= \frac{h}{B},
	&&\text{and}&
	\hat{u} &= \frac{u}{B^{1/2}}
\end{align}
so that, in the new variables, the barrier is of unit height. The initial conditions are therefore written as
\begin{align}
	\hat{h}(\hat{x},0) &= \hat{h}_i,
	&&\text{and}&
	\hat{u}(\hat{x},0) &= \hat{u}_i,
\end{align}
and the initial Froude number is $F_i \eqdef \hat{u}_i/\hat{h}_i^{1/2}$, and similarly for other subscripts. There are three types of solution possible, and all may be written as functions of $\hat{y} \eqdef \hat{x}/\hat{t}$ because the solutions are constant on the linear trajectories $\hat{x} \propto \hat{t}$. That is, at $\hat{t} = 0$ the variations in $\hat{h}$ and $\hat{u}$ are localised at $\hat{x}=0$, and then spread with the width increasing in proportion to $\hat{t}$ while maintaining their shape. 

The simplest solution type is a \emph{uniform state} which satisfies, for all $\hat{t}>0$,
\begin{subequations}\label{eqn:ot:boundRie_soln}
\begin{align}
	\hat{u} &= \hat{u}_i,
	&&\text{and}&
	\hat{h} &= \hat{h}_i,
	\label{eqn:ot:boundRie_soln_steady}
\end{align}
(see \cref{fig:ot:CritWall_BndRie_Prof1}(f,i) and \cref{fig:ot:CritWall_BndRie_Prof2}(d,i)). Alternatively the solution may take the form of a \emph{$\beta$-fan}, which is a simple wave across which $\alpha$ is constant and takes the value in the initial uniform state $\hat{\alpha} = \hat{u}_i + 2 \hat{h}_i^{1/2}$, while $\beta$ varies as a function of $\hat{y}$, that is
\begin{align}
	\hat{u} &= 
	\left\{\begin{array}{l r@{}l}
		\hat{u}_i,					& 						&\hat{y} \leq \hat{\mu}_i, 	\\
		(2\hat{y}+\hat{\alpha})/3,	&	\hat{\mu}_i \leq 	&\hat{y} \leq \hat{\mu}_b, 	\\
		\hat{u}_b,					&	\hat{\mu}_b \leq 	&\hat{y},
	\end{array}\right.
	&&\text{and}&
	\hat{h} &=
	\left\{\begin{array}{l r@{}l}
		\hat{h}_i,					& 						&\hat{y} \leq \hat{\mu}_i, 	\\
		(\hat{y}-\hat{\alpha})^2/9, &	\hat{\mu}_i \leq 	&\hat{y} \leq \hat{\mu}_b, 	\\
		\hat{h}_b,					&	\hat{\mu}_b \leq 	&\hat{y},
	\end{array}\right.
	\label{eqn:ot:boundRie_soln_fan}
\end{align}
where $\hat{\mu}_i \eqdef \hat{u}_i - \hat{h}_i^{1/2}$ and $\hat{\mu}_b \eqdef \hat{u}_b - \hat{h}_b^{1/2}$
(see \cref{fig:ot:CritWall_BndRie_Prof1}(a-e) and \cref{fig:ot:CritWall_BndRie_Prof2}(a-c)).
The final case is when the solution is a \emph{$\beta$-shock} between two uniform regions,
\begin{align}
	\hat{u} &= 
	\left\{\begin{array}{l r@{}l}
		\hat{u}_i,					& 	\hat{y} &< \hat{s}, 	\\
		\hat{u}_b,					&	\hat{s} &< \hat{y},
	\end{array}\right.
	&&\text{and}&
	\hat{h} &=
	\left\{\begin{array}{l r@{}l}
		\hat{h}_i,					&	\hat{y} &< \hat{s}, 	\\
		\hat{h}_b,					&	\hat{s} &< \hat{y},
	\end{array}\right.
	\quad\text{where}\quad
	\hat{s} \eqdef \frac{\hat{u}_b \hat{h}_b - \hat{u}_i \hat{h}_i}{\hat{h}_b - \hat{h}_i},
	\label{eqn:ot:boundRie_soln_shock}
\end{align}\end{subequations}
which must satisfy the shock condition \cref{eqn:ot:SW_shock_con_general_momentum} (see \cref{fig:ot:CritWall_BndRie_Prof1}(g,h) and \cref{fig:ot:CritWall_BndRie_Prof2}(e-h)).

\begin{figure}
	\centering
	\includegraphics{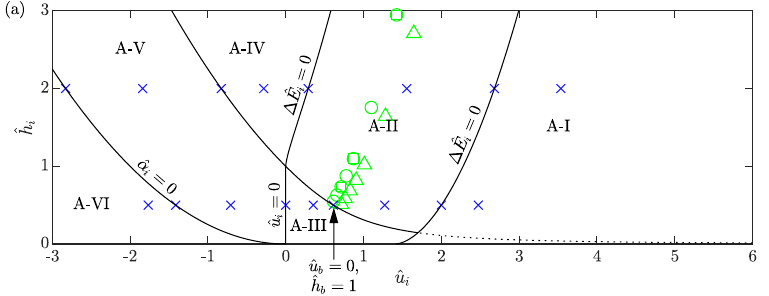}
	\includegraphics{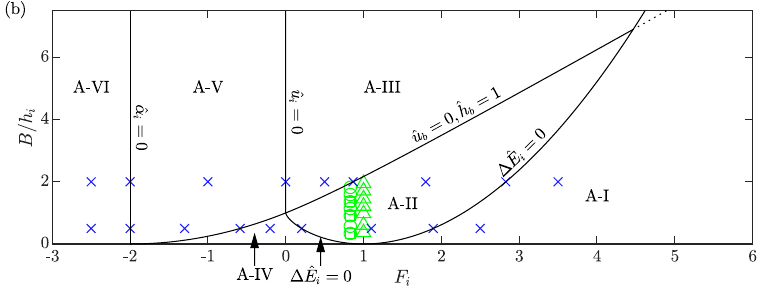}
	\caption{Classification of the overtopping flow in the boundary-Riemann problem (\cref{sec:overtop_close}) subject to the critical barrier boundary condition in (a) the $(\hat{u}_i,\hat{h}_i)$-plane and (b) the $(F_i,B/h_i)$-plane. The regions are labelled with the corresponding dynamical regime, and the dividing curves between them are labelled with their specifying equations. The dashed curve is the continuation of the dividing curve if supercriticality were not prioritised in the boundary condition. Points marked with $\times$ are the parameter values for \cref{fig:ot:CritWall_BndRie_Prof1,fig:ot:CritWall_BndRie_Prof2}, and those marked with $\square$/$\bigcirc$/$\triangle$ are the values for \cref{fig:ot:GonzalezJuez}. The overtopping regimes are: A-I, supercritical; A-II, subcritical with $\beta$-shock; A-III, blocked with $\beta$-shock; A-IV, subcritical with $\beta$-fan; A-V, blocked with $\beta$-fan; A-VI, dry.}
	\label{fig:ot:CritWall_BndRie}
\end{figure}

In the remainder of this section, for a given initial state we specify which of the outflow behaviours (supercritical, subcritical, or blocked) occur, and which of the three solution types is observed. This yields a division of the $(\hat{u}_i,\hat{h}_i)$-plane into 6 regimes as shown in \cref{fig:ot:CritWall_BndRie}. \Cref{fig:ot:CritWall_BndRie}(b) is similar to \cref{fig:ot:Baines_Davies_Regimes}, except that the axes are exchanged for consistency with, and the clarity of, later figures. The similarity is due to the upstream dynamics \emph{on} the dividing curves being the same for the two problems, though the dynamics \emph{within} some regimes differ (specifically A-IV, also A-V and A-VI are not present in \cref{fig:ot:Baines_Davies_Regimes}). To aid the construction of regimes we define $\Delta \hat{E}_i \eqdef \Delta E (\hat{u}_i,\hat{h}_i,1)$, which is the scaled energy discrepancy associated with the initial conditions, and $\Delta \hat{E}_b \eqdef \Delta E (\hat{u}_b,\hat{h}_b,1)$, which is that associated with the boundary values.

The first two regimes are easy to find:

\emph{Supercritical overtopping} (regime A-I in \cref{fig:ot:CritWall_BndRie}, see \cref{fig:ot:CritWall_BndRie_Prof1}(i) and  \cref{fig:ot:CritWall_BndRie_Prof2}(i)) occurs when $\hat{u}_i > \hat{h}_i^{1/2}$ and $\Delta \hat{E}_i > 0$, no boundary condition need be applied, resulting in a uniform state \cref{eqn:ot:boundRie_soln_steady}. The boundary of this regime is the curve $\Delta \hat{E}_i = 0$ (see \cref{fig:ot:CritWall_BndRie_Prof1}(h) and  \cref{fig:ot:CritWall_BndRie_Prof2}(h)).

A \emph{dry region} (regime A-VI in \cref{fig:ot:CritWall_BndRie}, see \cref{fig:ot:CritWall_BndRie_Prof1}(a) and  \cref{fig:ot:CritWall_BndRie_Prof2}(a)) is created adjacent to to the barrier when $\hat{u}_i < -2 \hat{h}_i^{1/2}$, and results in a $\beta$-fan \cref{eqn:ot:boundRie_soln_fan} with $\hat{h}_b = 0$, $\hat{\mu}_b = \hat{u}_i + 2 \hat{h}_i^{1/2}$. The boundary of this regime is the curve $\hat{u}_i = -2 \hat{h}_i^{1/2}$, equivalently $\alpha_i = 0$ (see \cref{fig:ot:CritWall_BndRie_Prof1}(b) and  \cref{fig:ot:CritWall_BndRie_Prof2}(b)) on which the depth vanishes precisely at the barrier.

The remaining solution regimes are constructed using the results in \cref{sec:overtop_app_Riemann}, which yield the following method. We identify the depth attained at the barrier for the boundary condition $\hat{u}_b = 0$, which we denote by $\hat{h}_{b0}$. When $\hat{u}_i = 0$ then $\hat{h}_{b0}=\hat{h}_i$; when $\hat{u}_i < 0$ the solution is a $\beta$-fan and because $\alpha$ is constant across the domain
\begin{subequations}\label{eqn:ot:boundrie_u0h1}\begin{gather}
	\hat{h}_{b0} = \ppar*{ \hat{h}_i^{1/2} + \frac{\hat{u}_i}{2} }^2;
\intertext{and when $\hat{u}_i > 0$ the solution is a $\beta$-shock and, by \cref{eqn:ot:SW_shock_con_general_momentum}, we seek the $\hat{h}_{b0}>\hat{h}_i$ solution to}
	(\hat{h}_{b0} + \hat{h}_i)(\hat{h}_{b0}-\hat{h}_i)^2 = 2 \hat{u}_i^2 \hat{h}_i \hat{h}_{b0}.
\end{gather}\end{subequations}
The boundary between the blocked and subcritical overtopping regimes occurs when $\hat{u}_b=0$ and $\hat{h}_b = 1$, thus from \cref{eqn:ot:boundrie_u0h1} the curve separating the regimes in the $(\hat{u}_i,\hat{h}_i)$-plane satisfies $\hat{h}_{b0}=1$. Solutions on this curve are plotted in \cref{fig:ot:CritWall_BndRie_Prof1}(d) and \cref{fig:ot:CritWall_BndRie_Prof2}(f).

\emph{Blocked flow with a $\beta$-fan} (regime A-V in \cref{fig:ot:CritWall_BndRie}, see \cref{fig:ot:CritWall_BndRie_Prof1}(c) and  \cref{fig:ot:CritWall_BndRie_Prof2}(c)) occurs when $\hat{h}_{b0} \leq 1$  and $\hat{u}_i < 0$, for which $\hat{u}_b = 0$, $\hat{h}_b = \hat{h}_{b0}$.

\emph{Blocked flow with a $\beta$-shock} (regime A-III in \cref{fig:ot:CritWall_BndRie}, see \cref{fig:ot:CritWall_BndRie_Prof2}(e)) occurs when $\hat{h}_{b0} \leq 1$ and $\hat{u}_i > 0$, for which $\hat{u}_b = 0$, $\hat{h}_b = \hat{h}_{b0}$.
Regimes A-V and A-III are separated by a dividing curve on which $\hat{h}_i \leq 1$ and $\hat{u}_i = 0$, corresponding to blocked flow in a uniform state (see \cref{fig:ot:CritWall_BndRie_Prof2}(d)).

\emph{Subcritical overtopping with a $\beta$-fan} (regime A-IV in \cref{fig:ot:CritWall_BndRie}, see \cref{fig:ot:CritWall_BndRie_Prof1}(e)) occurs when $\hat{h}_{b0} \geq 1$ and $\Delta \hat{E}_i > 0$, for which $\hat{u}_b$, $\hat{h}_b$ satisfy
\begin{subequations}\begin{align}
	&\begin{multicases}
		\Delta \hat{E}_b &= 0,	\\
		\hat{u}_b + 2 \hat{h}_b^{1/2} &= \hat{u}_i + 2 \hat{h}_i^{1/2},	\\
		1 \leq \hat{h}_b &\leq \min(\hat{h}_{b0},\hat{h}_i).
	\end{multicases}
\end{align}

\emph{Subcritical overtopping with a $\beta$-shock} (regime A-II in \cref{fig:ot:CritWall_BndRie}, see \cref{fig:ot:CritWall_BndRie_Prof1}(g) and  \cref{fig:ot:CritWall_BndRie_Prof2}(g)) occurs when $\hat{h}_{b0} \geq 1$ and $\Delta \hat{E}_i < 0$, for which $\hat{u}_b$, $\hat{h}_b$ satisfy
\begin{align}
	&\begin{multicases}
		\Delta \hat{E}_b &= 0,	\\
		(\hat{h}_b + \hat{h}_i)(\hat{h}_b - \hat{h}_i)^2 &= 2 (\hat{u}_b - \hat{u}_i)^2 \hat{h}_i \hat{h}_b,	\\
		\max(1,\hat{h}_i) \leq \hat{h}_b &\leq \hat{h}_{b0}.
	\end{multicases}
\end{align}\end{subequations}
Regimes A-IV and A-II are separated by a dividing curve on which $\hat{h}_i \geq 1$ and $\Delta \hat{E}_i = 0$, corresponding to subcritical outflow in a uniform state (see \cref{fig:ot:CritWall_BndRie_Prof1}(f)).

The division of the $(\hat{u}_i , \hat{h}_i)$-plane is plotted in \cref{fig:ot:CritWall_BndRie}(a), where black lines show division between the different dynamical regimes and are labelled with the equations that specify them. We note that across most of the domain the values at the barrier $\hat{u}_b$, $\hat{h}_b$ vary continuously as functions of $\hat{u}_i$, $\hat{h}_i$, and so states on the dividing curves can be constructed as a limit from either side. The only exception is across the boundary of the supercritical regime, where $\hat{u}_b$ and $\hat{h}_b$ are discontinuous as functions of $\hat{u}_i$, $\hat{h}_i$.  By our choice of boundary condition in \cref{sec:overtop_PF}, the solutions on this curve are the limit from the subcritical side (see \cref{fig:ot:CritWall_BndRie_Prof1}(h) and  \cref{fig:ot:CritWall_BndRie_Prof2}(h)).

It is also informative to view the regimes in the plane where velocity is measured relative to the local wave-speed, $F_i \eqdef {u_i}/{h_i^{1/2}} = {\hat{u}_i}/{\hat{h}_i^{1/2}} = \eval{{\hat{u}_l}/{\hat{h}_l^{1/2}}}_{\hat{t}=0}$, and barrier height relative the height of the incident current, $B/h_i = \hat{h}_i^{-1} = \eval{\hat{h}_l^{-1}}_{\hat{t}=0}$, see \cref{fig:ot:CritWall_BndRie}(b). This is particularly useful when the oncoming current current has a specified Froude number $\Fro$ at its front, only a single vertical slice through the $(F_i , \hat{h}_i^{-1})$-plane is required. For this purpose, it is useful to know that the curve $\Delta \hat{E}_i = 0$ intersects with the $\hat{u}_b = 0$, $\hat{h}_b = 1$ curve when $(F_i , \hat{h}_i^{-1}) \in\pbrc{ (0,1) , (4.47,6.90) }$. Thus, for $0 < \Fro < 1$, a barrier slightly taller than the current prevents overtopping, and otherwise there is subcritical overtopping. For $1  < \Fro < 4.47$, relatively tall barriers prevent overtopping, moderate height barrier cause subcritical overtopping, and relatively short barriers cause supercritical overtopping. For $\Fro > 4.47$, very tall barriers prevent overtopping, and all others cause supercritical overtopping.

%% file: Sections/Collision.tex
\section{Unsteady overtopping of a barrier} \label{sec:overtop_collision_extreme}

Using the behaviour local to a barrier governed by the critical barrier boundary condition, we investigate the unsteady dynamics of the system formulated in \cref{sec:overtop_PF}. In this section we consider the collision of the gravity current with the barrier in two extreme cases: first when the barrier is sufficiently close to the release that there is no effect of the back-wall of the lock prior to collision; and second when the barrier is far enough from the release that the current takes on similarity form prior to collision. As time progresses the outflow may transition between supercritical overtopping, subcritical overtopping, and blocked flow. In what follows, we elucidate both the initial mode at collision and how the flow transitions between modes over time.

\subsection{Barrier close to initial release}\label{sec:overtop_collision_close}

We assume that the barrier is sufficiently close that the presence of the back-wall is not felt by the front of the current prior to impact. At the instant of release, the influence of the lock's removal creates a wave that travels to the back-wall, which is then reflected forward towards the front of the gravity current \citep{ar_Hogg_2006}. We neglect the influence of this forward propagating reflected wave in this section, which requires that the barrier is sufficiently close to the initial release (see \cref{sec:overtop_simulation}). Formally we let $L = 1+ \tilde{\delta}$ for some $\epsilon \ll \tilde{\delta} \ll 1$, and define $\tilde{x} = \frac*{(x-1)}{\tilde{\delta}}$, $\tilde{t} = \frac*{t}{\tilde{\delta}}$ so that, as $\tilde{\delta} \rightarrow 0$, the initial release occupies $\tilde{x} < 0$ and the barrier is at $\tilde{x} = 1$. At the instant of release the solution is a $\beta$-fan originating at $\tilde{x}=0$ that joins two uniform regions, and given explicitly by
\begin{subequations}\label{eqn:ot:lockrel_early_soln}
\begin{gather} \label{eqn:ot:lockrel_early_soln_uh}
	\begin{aligned}
		u &= \frac{2 F}{F + 2},			
		&\qquad&&
		h &= \ppar*{ \frac{2}{F + 2} }^2
	\end{aligned}
\shortintertext{where}
	F =
	\begin{multicasesd}[3]
		&0 															&& \text{for } & 									&	\frac{\tilde{x}}{\tilde{t}} \leq -1,	\\
		&\frac{2 + 2 (\tilde{x}/\tilde{t})}{2 - (\tilde{x}/\tilde{t})}	&& \text{for } & 						-1 \leq 	&	\frac{\tilde{x}}{\tilde{t}} \leq 2 \frac{\Fro - 1}{\Fro + 2},	\\
		&Fr															&& \text{for } & 2 \frac{\Fro - 1}{\Fro + 2} \leq	&	\frac{\tilde{x}}{\tilde{t}} \leq \frac{2 \Fro}{\Fro + 2}
	\end{multicasesd}
	\label{eqn:ot:lockrel_early_soln_Fr}
\end{gather}
\end{subequations}
is the local Froude number. We express the solution in terms of $F$ rather than directly in terms of $\tilde{x}/\tilde{t}$ \citep[\eg][]{bk_Ungarish_GCIv2} for analytical simplicity in what follows. Note that $F$ increases monotonically with $\tilde{x}/\tilde{t}$ from $0$ to $\Fro$ across the $\beta$-fan (the middle case in \cref{eqn:ot:lockrel_early_soln_Fr}).

\begin{figure}
	\centering
	\includegraphics{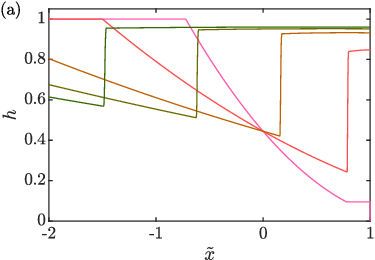}
	\includegraphics{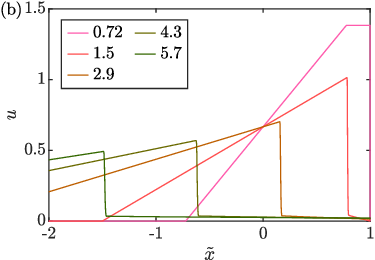}
	\includegraphics{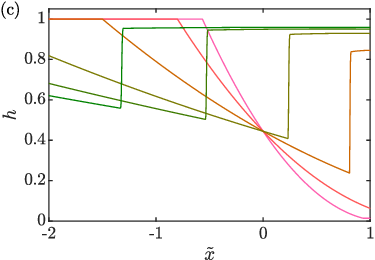}
	\includegraphics{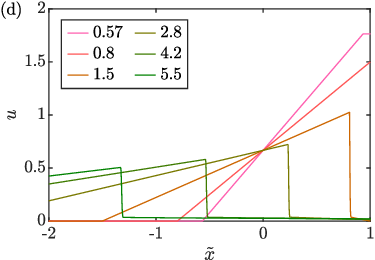}
	\includegraphics{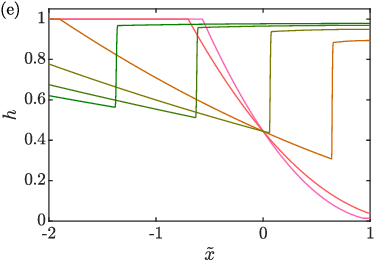}
	\includegraphics{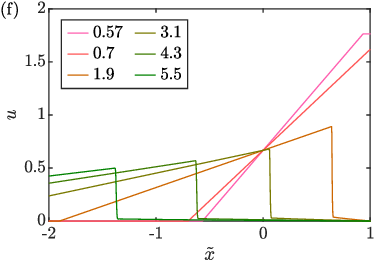}
	\includegraphics{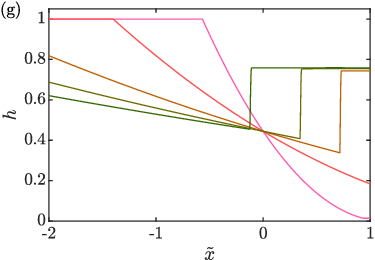}
	\includegraphics{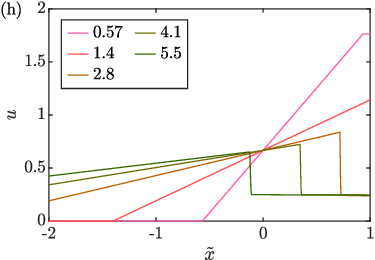}
	\caption{The depth, $h$, and velocity, $u$, as functions of the scaled spatial coordinate, $\tilde{x}$, at various instances of time. The pairs of plots (a,b), (c,d) (e,f), and (g,h) are at parameter values of $B$ and $\Fro$ marked on \cref{fig:ot:collision_close}. The times are indicated by differing colours, with values given by the legends in (b), (d), (f), and (h). In (a,b) $\Fro = 4.5$, $B = 0.85$, the times correspond to the instant of collision (blocked flow), a time just prior to critical overtopping, and three later times. In (c,d) $\Fro = 15$, $B = 0.85$, the times correspond to the instant of collision (supercritical overtopping), a time just prior to overtopping ceasing, a time just prior to critical overtopping starting, and four later times. In (e,f) $\Fro = 15$, $B = 1.05$, the times correspond to the instant of collision (supercritical overtopping), a time just prior to overtopping ceasing, and four later times. In (g,h) $\Fro = 15$, $B = 0.3$, the times correspond to the instant of collision (supercritical overtopping), a time just prior to the supercritical overtopping transitioning to subcritical overtopping, and three later times.}
	\label{fig:ot:CritWall_Inf_Profiles}
\end{figure}

The solution \cref{eqn:ot:lockrel_early_soln} is valid until the front reaches the barrier, which occurs at time $\tilde{t} = \frac*{(\Fro + 2)}{(2 \Fro)}$. To illustrate our analysis of the subsequent dynamics we use plots produced from numerical simulation (\cref{fig:ot:CritWall_Inf_Profiles}). These simulations were performed on a mesh of $1000$ cells on $-2 \leq \tilde{x} \leq 1$ initiated at the instant of collision, with a non-reflecting boundary condition imposed at $\tilde{x} = -2$ so that this boundary does not influence the ensuing dynamics.

\begin{figure}
	\centering
	\includegraphics{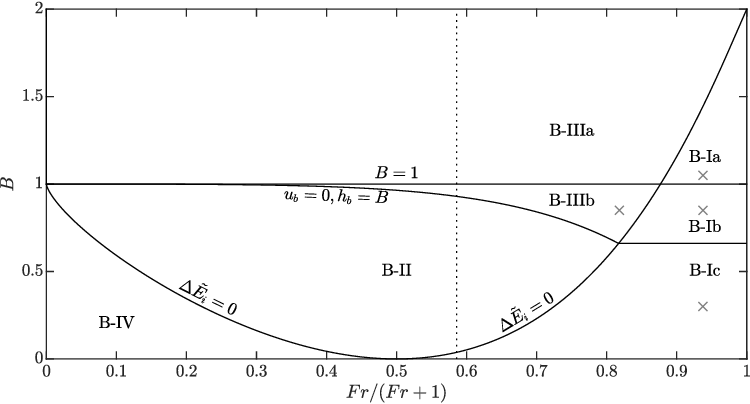}
	\caption{Classification of the overtopping flow when a gravity current collides with a barrier that is sufficiently close to the initial release, with labels for each regime (\cref{sec:overtop_collision_close}). The boundary curves (solid) are labelled with the equation solved to compute them. The parameter values used in \cref{fig:ot:CritWall_Inf_Profiles} are marked by $\times$. On the abscissa, we employ a transformation so that $\Fro=0$ at the left and $\Fro \to \infty$ at the right. The vertical dotted line represents $\Fro=\sqrt{2}$, a Boussinesq current. The overtopping regimes are: B-I, supercritical transitioning to B-Ia blocked, B-Ib blocked then subcritical, B-Ic subcritical; B-II, subcritical; B-III, blocked, in B-IIIb transitioning to subcritical; B-IV subcritical, but generating a $\beta$-fan at the barrier rather than a $\beta$-shock.}
	\label{fig:ot:collision_close}
\end{figure}

The initial interaction with the barrier is now discussed. We emphasise that our model does not include mixing between the current and ambient fluid, which may occur during the initial collision between Boussinesq flows and relatively deep barriers. We note that the dilute fluid generated by this interaction will largely be transported downstream out of the domain, and should not affect the subsequent motion. The interaction corresponds to the boundary-Riemann problem solved in \cref{sec:overtop_close}. In particular, since $\Fro > 0$, we find that four initial behaviours may occur, depending on $\Fro$ and the relative height of the barrier: supercritical overtopping (B-Ia, B-Ib, B-Ic); subcritical overtopping with a  $\beta$-shock (B-II) or a $\beta$-fan (B-IV) generated at the barrier; and blocked flow (B-IIIa, B-IIIb), see \cref{fig:ot:collision_close}. These are divided by two curves. Firstly  $\Delta \tilde{E}_i \eqdef \Delta \tilde{E} (\Fro,B) = 0$ is the boundary of supercritical overtopping, with $\Delta \tilde{E}_i$ the energy discrepancy when the front arrives at the barrier. The energy discrepancy across the domain, denoted $\Delta \tilde{E}$, is derived by substituting \cref{eqn:ot:lockrel_early_soln_uh} into \cref{eqn:ot:energy_discrep}, yielding
\begin{equation}\label{eqn:ot:lockrel_early_soln_energy}
	\Delta \tilde{E}(F,B) \eqdef \Delta E(u,h,B) = \ppar*{\frac{2}{F+2}}^2 \ppar*{\textfrac{1}{2} F^2 + 1 - \textfrac{3}{2} F^{2/3}} - B.
\end{equation}
Secondly, $u_b = 0$, $h_b = B$, which by the shock conditions \cref{eqn:ot:SW_shock_con_general} imply $\tilde{S}(\Fro,B) = 0$, where
\begin{equation}\label{eqn:ot:lockrel_early_soln_u0hB}
	\tilde{S}(F,B) \eqdef \pbrk*{ B + \ppar*{\frac{2}{F+2}}^2} \pbrk*{ B - \ppar*{\frac{2}{F+2}}^2}^2 - 2 B F^2 \ppar*{\frac{2}{F+2}}^4
\end{equation}
divides subcritical and blocked flow. In what follows we analyse the subsequent motion that occurs and we document the results in terms of the dimensionless height of the barrier $B$ and the Froude number at the front of the flow $\Fro$, plotting the dynamics in \cref{fig:ot:CritWall_Inf_Profiles} and including the regimes in \cref{fig:ot:collision_close}.

If, at the instant of collision, the flow is blocked, then initially there will be a $\beta$-shock that travels from the barrier with a constant velocity until it interacts with the $\beta$-fan from the initial conditions. From this time on, the region of fluid between the shock and barrier progressively deepens, eventually attaining unit depth (\cf \citealt{ar_Skevington_F002_SemiInf_Wall}). Therefore, for barriers of height less than $1$ the fluid eventually overtops (\cref{fig:ot:CritWall_Inf_Profiles}(a,b)), which we call regime B-IIIb, whereas for a barrier of height greater than $1$ the fluid never overtops, regime B-IIIa.

Alternatively, if at the instant of collision the fluid supercriticality overtops then no shock is generated and instead the solution for early $\tilde{t}$ is valid until $\Delta E_b = 0$ at the barrier. Thus the supercritical outflow will cease while the $\beta$-fan from the initial conditions is adjacent to the barrier. Specifically it will cease when $\Delta \tilde{E}(F,B)=0$ (for a given $B$), which can be solved to find $F$ which is the Froude number adjacent to the barrier. At this instant we must solve another boundary-Riemann problem, but it is substantively the same as the one solved above for the initial interaction, with $\alpha=2$ and the Froude number now being $F$. Consequently, the division between transition to subcritical overtopping (B-Ic, see \cref{fig:ot:CritWall_Inf_Profiles}(g,h)) and transition to blocked flow (B-Ia, B-Ib) is a line of constant $B$ which can be found by solving $\Delta \tilde{E}(F,B)=\tilde{S}(F,B)=0$, yielding $(F,B) = (4.47,0.661)$ (3 \sf). As for flows that initially do not overtop (B-IIIa, B-IIIb), flows that transition from supercritical overtopping to blocked flow can subsequently transition to subcritical overtopping (B-Ib, see \cref{fig:ot:CritWall_Inf_Profiles}(c,d)) if $B<1$, and do not (B-Ia, see \cref{fig:ot:CritWall_Inf_Profiles}(e,f)) if $B>1$. On the dividing line between these regimes the supercritical overtopping ceases when $\Delta \tilde{E}(F,1)=0$, thus $F = 7.12$ (3\sf). We note that the given values of $F$ correspond to the values of $\Fro$ on the bounding curve of supercritical overtopping $\Delta E_i=0$, and therefore mark the locations at which the dividing curves meet.

We note that as $\tilde{t} \rightarrow \infty$ the solution limits to the same solution as the boundary-Riemann problem with $\hat{h}_i = 1/B$, $\hat{u}_i = 0$ (\cref{sec:overtop_close}). Therefore with $B<1$ we find subcritical overtopping, whereas with $B>1$ the flow is blocked at late times. For this reason, once these modes of overtopping have been reached, we expect no further changes.

The classification of dynamics is shown in \cref{fig:ot:collision_close}. One surprising result is that for $\Fro > 7.12$ the initial collision can \emph{only} produce supercritical overtopping or blocked flow, and that supercritical is possible for $B>1$. That is to say, fluid can overtop a barrier taller than the initial release, and for $\Fro \rightarrow \infty$ it can overtop a barrier twice the height of the release, as a consequence of energy conservation.

We emphasise that these results are valid before the finite extent of the release influences the overtopping through the arrival at the barrier or shock of the characteristic reflected from the back-wall (formally $\tilde{\delta} \rightarrow 0$). The effects of a finite extent will modify the boundaries between the regimes in a way that depends on $L$. 

\subsection{Barrier far from initial release}\label{sec:overtop_collision_far}

\begin{figure}
 	\centering
 	\includegraphics{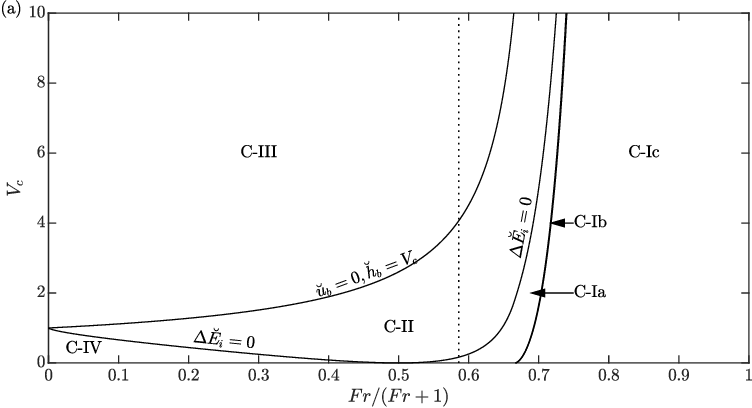}
 	\includegraphics{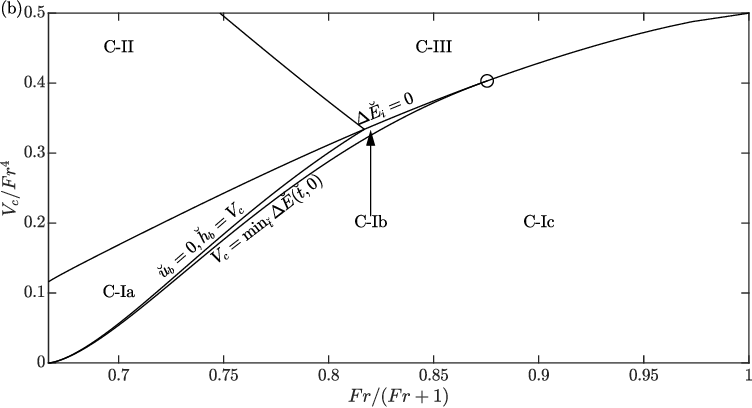}
 	\caption{Classification of the overtopping flow for a gravity current in similarity form (\cref{sec:overtop_collision_far}); (b) is the same as (a) except for a rescaling of the axes. Each regime is labelled, and the boundary curves (solid) are labelled with the equation solved to compute them. The circle marks the intersection of two dividing curves at $\Fro = 7.03$, $V_c = 984$.  On the abscissa, we employ a transformation so that $\Fro=0$ at the left of (a) and $\Fro \to \infty$ at the right, while in (b) we show $2 < \Fro < \infty$.  The vertical dotted line in (a) represents $\Fro=\sqrt{2}$, a Boussinesq current. The overtopping regimes are: C-I, supercritical transitioning to C-Ia subcritical then blocked, C-Ib blocked, C-Ic dry; C-II, subcritical then blocked; C-III blocked; C-IV subcritical, but generating a $\beta$-fan at the barrier rather than a $\beta$-shock, transitioning to blocked.}
 	\label{fig:ot:collision_far}
\end{figure}

We now assume that the barrier is sufficiently far from the initial release that the current is in similarity form when it reaches the barrier, which is the late time limit for all finite values of the Froude number, $\Fro$, at the front of the current \citep{ar_Gratton_1994}. We express the similarity solution in terms of the variables
\begin{subequations}
\begin{align}
	\breve{x} &= \frac{x}{L},	&
	\breve{t} &= \frac{t-t_0}{L^{3/2}}, &
	\xi &= \frac{K \breve{x}}{\breve{t}^{2/3}},
\end{align}
where $t_0$ is some arbitrary time offset and
\begin{align}
	K &= \ppar*{\frac{2}{3}}^{\frac*{2}{3}} \ppar*{ \frac{1}{\Fro^2} + \frac{\xi_D^3 - 1}{6} }^{\frac*{1}{3}},	&
	\xi_D &= \max\ppar*{ 1 - \frac{4}{\Fro^2} , 0}^{1/2}.
\end{align}
The similarity solution is, in terms of the variables $\breve{u} \eqdef L^{1/2} u$, $\breve{h} \eqdef L h$,
\begin{align}
	\breve{u} &= \frac{2 \xi}{3 K \breve{t}^{\frac*{1}{3}}},	&
	\breve{h} &= \frac{1}{9 K^2 \breve{t}^{\frac*{2}{3}}} \ppar*{ \xi^2 + \frac{4}{\Fro^2} - 1 }	&
	\text{for}&&
	\xi_D &\leq \xi \leq 1,
	\\
	&&
	\breve{h} &= 0	&
	\text{for}&&
	0 &\leq \xi < \xi_D.
\end{align}
\end{subequations}
In this rescaled system the remaining parameters are $V_c = BL$ (the confined volume \cref{eqn:confined_vol}) and $\Fro$ (the front condition \cref{eqn:front_condition}), and we determine the dynamical regimes after collision in terms of them. The similarity solution is valid until the front reaches the barrier, which occurs at time $\breve{t} = K^{\frac*{3}{2}}$. At this time a boundary-Riemann problem must be solved at $\breve{x} = 1$ to determine the dynamics immediately after the collision (see \cref{sec:overtop_close}). This initial local interaction divides parameter space  into four regimes corresponding to: blocked flow (C-III); subcritical overtopping with a $\beta$-shock (C-II) or a $\beta$-fan (C-IV) generated at the barrier; and supercritical overtopping (C-Ia, C-Ib, C-Ic) (see \cref{fig:ot:collision_far}). These regimes are divided by two curves. Firstly, the curve $\Delta \breve{E}_i \eqdef \Delta \breve{E} (K^{3/2},V_c) = 0$ (energy discrepancy at impact), where 
\begin{align}\label{eqn:ot:lockrel_far_soln_energy}
 	\Delta \breve{E} (\breve{t},V_c) 
 	&\eqdef \left. \Delta E(\breve{u},\breve{h},V_c) \right|_{\breve{x} = 1} \\ \notag
 	&= \frac{1}{9 K^2 \breve{t}^{2/3}} \pbrk*{ \frac{3 K^2}{\breve{t}^{4/3}} + \frac{4}{\Fro^2} - 1 - \frac{3 K^{2/3}}{2^{1/3} \breve{t}^{4/9}} \ppar*{ \frac{K^2}{\breve{t}^{4/3}} + \frac{4}{\Fro^2} - 1 }^{2/3} } - V_c,
\end{align}
(energy discrepancy computed at $\breve{x} = 1$ across $\xi_D \leq \xi \leq 1$) is the boundary of supercritical overtopping. Secondly, $\breve{u}_b = 0$, $\breve{h}_b = V_c$, which by the shock conditions \cref{eqn:ot:SW_shock_con_general} imply $\breve{S}(K^{3/2},V_c) = 0$, where
\begin{multline}\label{eqn:ot:lockrel_far_soln_u0hB}
	\breve{S} (\breve{t},V_c) 
	= \pbrk*{ V_c + \frac{1}{9 K^2 \breve{t}^{2/3}} \ppar*{\frac{K^2}{\breve{t}^{4/3}} + \frac{4}{\Fro^2} - 1} } \pbrk*{ V_c - \frac{1}{9 K^2 \breve{t}^{2/3}} \ppar*{\frac{K^2}{\breve{t}^{4/3}} + \frac{4}{\Fro^2} - 1} }^2	\\
	- \frac{2^3 V_c}{3^4 K^2 \breve{t}^{8/3}} \ppar*{\frac{K^2}{\breve{t}^{4/3}} + \frac{4}{\Fro^2} - 1},
\end{multline}
is the boundary between subcritical overtopping and blocking. In what follows we analyse the subsequent transitions of overtopping behaviour that occur within each of the regimes.
 
If the fluid is unable to supercritically overtop at the instant of collision (\ie, blocked flow or subcritical overtopping) then two cases are possible. One case is when a $\beta$-fan is generated at the subcritically overtopping barrier (C-IV), so that the depth at the barrier $\breve{h}_b$ is smaller than the depth that arrived at the front. This continues until $\breve{h}_b$ reduces to $V_c$ at which time the flow is blocked. The other case is when an upstream propagating shock is formed. We find from simulation that, as time passes, the fluid depth between the shock and wall decreases. This can be rationalised as follows. Initially, the distance between the shock and barrier is small, and therefore the dynamics in this regime is quasi-static and has constant velocity throughout. Using the perturbed shock condition \cref{eqn:ot:perturbshock_uh}, and the fact that upstream of the shock both $u$ and $h$ are decreasing, we deduce that $h$ downstream of the shock must decrease also. At later times the fluid is not quasi-static, and therefore we make use of the expressions involving the invariants, \cref{eqn:ot:perturbshock_ab}, remembering that the shock sets the changes in $\alpha$ immediately downstream of the shock, these changes are then reflected off the barrier resulting in changes of opposite sign to $\beta$. The large coefficient of the perturbations to $\alpha$ suggests that $\alpha$ is decreasing and, therefore, $\beta$ is increasing, corresponding to a decreasing depth. Because the fluid adjacent to the barrier is becoming shallower, we deduce that the fluid will cease its initial overtopping (if any), and not overtop again. Thus, in regime C-III, no fluid makes it over the barrier, and in regime C-II the subcritical overtopping ceases after finite time.

For supercritical outflow we find that there are three possible types of subsequent motion. Firstly, it is possible that all of the gravity current may overtop the barrier. If $\min_{\breve{t}} \Delta \breve{E}(\breve{t},V_c) \geq 0$ (equivalent to $\min_{\breve{t}} \Delta \breve{E}(\breve{t},0) \geq V_c$) the energy of fluid reaching the barrier is sufficient to allow supercritical overtopping for all time, thus all of the fluid flows out of the domain supercritically (C-Ic). We note that for $\Fro \geq 7.03$ (3\sf) the time of minimal $\Delta \breve{E} (\breve{t},0)$ is at the instant of collision, $\breve{t} = K^{3/2}$ (at $\Fro=7.03$, $\Delta \breve{E}(K^{3/2},0) = 984$), thus if there is supercritical overtopping then it will certainly dry the domain. For supercritical overtopping that is not a part of C-Ic, we identify the time at which $\Delta \breve{E}(\breve{t},V_c) = 0$, and at this instant the fluid can either transition to subcritical overtopping (C-Ia) or blocked flow (C-Ib), following the boundary-Riemann problem at that instant. The dividing curve is when $\breve{h} = V_c$ and $\breve{u} = 0$ to the right of the $\beta$-shock generated at the barrier, and to find it we solve $\Delta \breve{E}(\breve{t},V_c) = \breve{S}(\breve{t},V_c) = 0$ as a coupled system for $\breve{t}$ and $V_c$. This curve will meet the corresponding dividing curve for subcritical outflow at the boundary of the supercritical regime, that is $\Delta \breve{E}(K^{3/2},V_c) = \breve{S}(K^{3/2},V_c) = 0$ which has solution $\Fro = 4.47$, $V_c=133$. In addition, it can be shown that the $V_c = 0$ solution is found at $\Fro = 2$. Note that, as in region C-II, the subcritical outflow of C-Ia will eventually transition to no-overtopping.

Classification of the dynamics is shown in \cref{fig:ot:collision_far}. For $\Fro \gg 1$ it is found that the current is able to surmount barriers that are much taller than the average depth of the current, that is $V_c$ is able to be very large and still permit outflow. This is because the current has most of its volume at the front, moving forwards with very high momentum. For more modest $\Fro$ this effect is weakened.

%% file: Sections/VolumeOutCritWall.tex
\section{Numerical evaluation of fluid outflow over a critical barrier}\label{sec:overtop_simulation}

Outside of the extremal cases discussed in \cref{sec:overtop_close,sec:overtop_collision_extreme}, developing a comprehensive understanding of the fluid flow dynamics is challenging to do analytically; instead we proceed with a numerical study of the dynamics. We compute two illustrative cases: $\Fro = \sqrt{2}$ corresponding to Boussinesq currents under a deep ambient \citep{ar_Benjamin_1968}; and $\Fro \rightarrow \infty$ when the ambient is of a substantially lower density than the current.

\subsection{Boussinesq gravity current: $\Fro = \sqrt{2}$}

We begin by establishing how tall the barrier must be to prevent outflow. To find this we run simulations imposing an insurmountable barrier at $x=L$ (that is impose the condition $u=0$ when the current reaches the barrier, which is equivalent to the limit $B \rightarrow \infty$), the shortest barrier required to prevent outflow will be equal to the maximum depth of fluid at the barrier over all time. 

\begin{figure}
	\centering
	\includegraphics{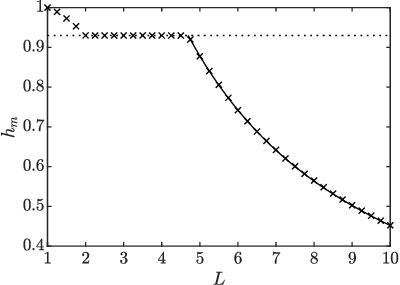}
	\caption{The maximum depth of fluid at the barrier, $h_m$, as a function of the location of the barrier, $L$, for a (fully reflected) Boussinesq current ($\Fro = \sqrt{2}$). Crosses show numerical data; dotted line the maximum depth when the current is uniform at the front; and solid line the exact solution \citep{ar_Hogg_2006}.}
	\label{fig:ot:InfWall_MaxDepth}
\end{figure}

The variation of the maximum depth of fluid, $h_m$, with downstream distance to the barrier, $L$, arises from the interaction of two key flow features: the upstream moving shock that is generated by the arrival of the flow at the barrier, and the trajectory of the `reflected’ $\alpha$-characteristic from the back-wall of the lock $(x=0)$ in response to the arrival of the rearmost $\beta$-characteristic formed by the removal of the lock-gate. It is this latter characteristic that `communicates’ the finite extent of the flow. If the barrier is sufficiently distant from the lock, then the reflected $\alpha$-characteristic catches up with the front of the motion before the gravity current reaches the barrier; however for a closer barrier this characteristic intersects the shock.  In \cref{fig:ot:InfWall_MaxDepth}, we plot the dependence of the maximum depth, $h_m$, on $L$ and observe that there are two values of $L$ at which the behaviour changes.  In what follows we demonstrate how to evaluate these transitions.

In \cref{sec:overtop_collision_close} we showed that the initial motion of the gravity current features a uniform region, leading a $\beta$-fan in which the depth and velocity vary (see \cref{eqn:ot:lockrel_early_soln}).  On reflection from the barrier, the frontal uniform region develops a uniform reflection in a region adjacent to the barrier within which the fluid is motionless and the shock moves steadily upstream (see Hogg \& Skevington 2021).  If the barrier is very close to the initial release $(L-1\ll 1)$ then the fluid progressively deepens at the barrier after the uniform state (reaching the asymptotic values $h_m\to 1$ as $L\to 1$).  The deepening stops when the reflected $\alpha$-characteristic emanating from $x=0$ intersects the shock and then propagates further to the barrier.  The time between the first arrival of the front of the gravity current and this $\alpha$-characteristic is reduced with increasing $L$ in this regime, and so the maximum depth, $h_m$, also reduces with $L$ (see \cref{fig:ot:InfWall_MaxDepth} with $1<L<2.02$).

\begin{figure}
	\centering
	\includegraphics{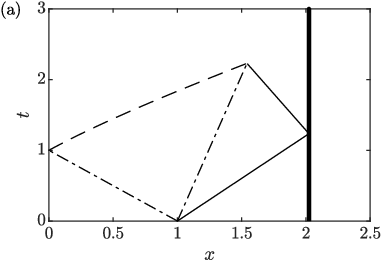}
	\includegraphics{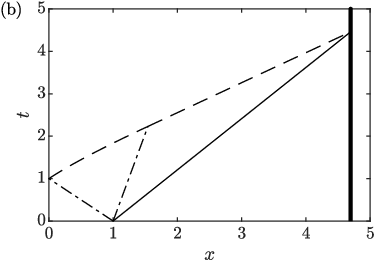}
	\caption{Plots of the characteristic trajectories for the collision of a gravity current with an insurmountable barrier for (a) $L = L_A$ and (b) $L=L_B$, both for $\Fro = \sqrt{2}$. The dot dash lines are $\beta$-characteristics, dashed lines are $\alpha$-characteristics, solid lines are the front and shock, and bold lines are the location of the barrier.}
	\label{fig:ot:InfWall_Lrange}
\end{figure}

A change in dependence of $h_m$ with $L$ occurs if the $\alpha$-characteristic intersects the shock when the fluid beyond it is still in a uniform state.  From \cref{eqn:ot:lockrel_early_soln}, the depth and velocity of the front of the gravity current are given by $u_f=2Fr/(Fr+2)$ and $h_f=4/(Fr+2)^2$.  On reflection from the barrier the fluid is motionless and of depth $h_b$, which by \cref{eqn:ot:SW_shock_con_general} satisfies
\begin{equation}
	2u_f^2 h_f h_b=(h_f-h_b)^2(h_f+h_b).
\end{equation}
For example, when $Fr=\sqrt{2}$ we find $h_b = 0.930$ (3 \sf). The speed of the shock is $s=u_fh_f/(h_f-h_b)$ when the gravity current front first arrives at the barrier at $t=(L-1)/u_f$. The position of the shock is therefore given by
\begin{equation}
	x_s = L + \frac{u_fh_f}{h_f-h_b} \ppar*{ t-\frac{L-1}{u_f} }.\label{eqn:ot:maxdepth_shock_pos}
\end{equation}
The reflected $\alpha$-characteristic meets the uniform region when
\begin{equation}
	x=1+(Fr-1)\left(\frac{Fr+2}{2}\right)^{1/2}\qquad\hbox{and}\qquad t=\left(\frac{Fr+2}{2}\right)^{3/2}\label{eqn:ot:maxdepth_char_pos}
\end{equation}
\citep[see][]{ar_Hogg_2006}. Thus the distance from the lock, $L=L_A$, at which the change of behaviour occurs is determined by substituting \eqref{eqn:ot:maxdepth_char_pos} into \eqref{eqn:ot:maxdepth_shock_pos}. For $Fr=\sqrt{2}$, we find $L_A=2.02$ (3 \sf; see \cref{fig:ot:InfWall_Lrange}(a)).
	
For $L_A<L<L_B$, the $\alpha$-characteristic intersects the shock while the reflected motion is uniform, and the maximum depth, $h_m$, is constant and equal to $h_b$ at the instant of collision. The next change of behaviour is at $L=L_B$ when the characteristic catches up with the front at the instant of collision (\cref{fig:ot:InfWall_Lrange}(b)). Using the results in \cite{ar_Hogg_2006}, this occurs for
\begin{equation}
	L_B = 1 + 2Fr \ppar*{\frac{\Fro + 2}{2}}^{\mathrlap{1/2}},
\end{equation}
and thus $L_B=4.70$ (3 \sf) for $Fr=\sqrt{2}$.
	
For a more distant barrier, $L>L_B$, the fluid depth at the front of the current begins to diminish before it reaches the barrier.  In this scenario, the maximum depth at the barrier, $h_m$, is determined by the depth of the arriving current, which may be evaluated directly from \cite{ar_Hogg_2006}.  The results are plotted in \cref{fig:ot:InfWall_MaxDepth}, which shows how $h_m$ diminishes with $L$.

For barriers which permit overflow we may expect to see some supercritical overtopping, but for $\Fro = \sqrt{2}$ this has been found to only occur for very short barriers indeed. Using the results of \cref{sec:overtop_close} and the early time solution \cref{eqn:ot:lockrel_early_soln} we find that supercritical overtopping only occurs for $B \lesssim 0.038$. We do not consider $B$ this small, and instead focus on barriers which cause subcritical overtopping or blocked flow.

\begin{figure}
	\centering
	\includegraphics{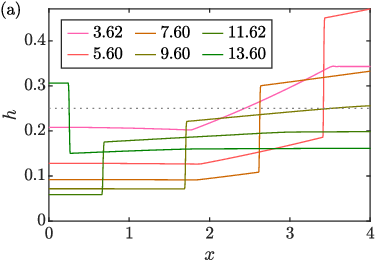}
	\includegraphics{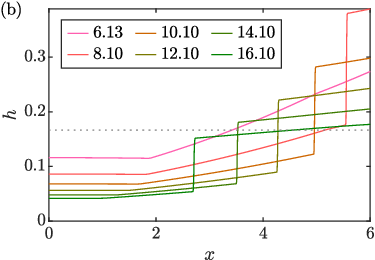}
	\includegraphics{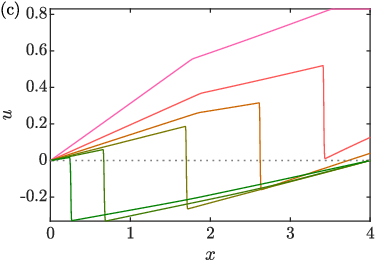}
	\includegraphics{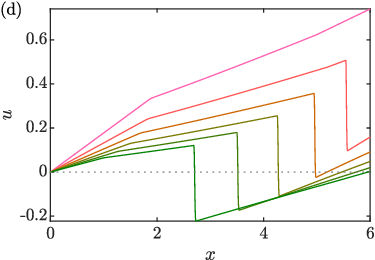}
	\includegraphics{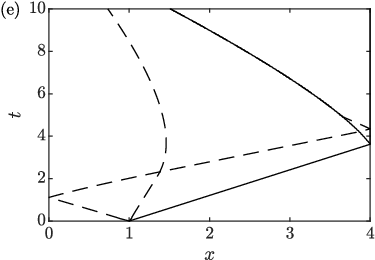}
	\includegraphics{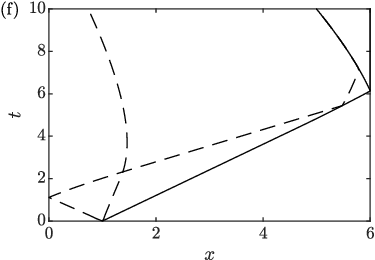}
	\includegraphics{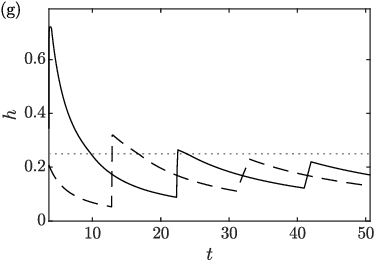}
	\includegraphics{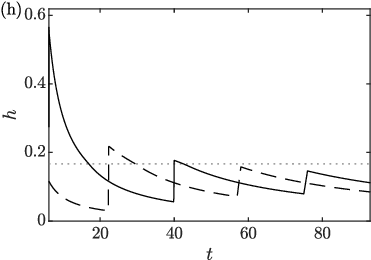}
	\caption{(a,b) The depth, $h$, as a function of $x$ at various instances of time as indicated in the legend, with $h=B$ in grey dotted. The first time plotted is when the current reaches the barrier. (c,d) The velocity, $u$, as a function of $x$ at the times in the legend of (a,b) respectively, with $u=0$ in grey dotted. (e,f) The characteristic plane, with gradient discontinuities in dashed lines, and the front and shock locations in solid lines. (g,h) The depth at $x=0$ (dashed) and $x=L$ (solid), with $h=B$ in grey dotted. Figures (a,c,e,g) are for $L = 4$, $B = 1/4$, while (b,d,f,h) are for $L=6$, $B=1/6$. }
	\label{fig:ot:CritWall_Fin_Profiles}
\end{figure}

In \cref{fig:ot:CritWall_Fin_Profiles} we plot the results from our numerical computations for $L=4$, $B=1/4$ and $L=6$, $B=1/6$. On reaching the barrier, both flows overtop subcritically but the nature of the interaction at early times differs between the two cases. Following the analysis above for insurmountable barriers, when $L<4.70$, the overtopping is not initially affected by the finite extent of the release (through the reflection of the characteristic from the back-wall, \cref{fig:ot:CritWall_Fin_Profiles}(e)), whereas for $L>4.70$ it is (\cref{fig:ot:CritWall_Fin_Profiles}(f)). This means that the former case exhibits a short period during which the overtopping flux is constant (\cref{fig:ot:CritWall_Fin_Profiles}(a,c)) and thereafter it varies temporally.

After the initial overtopping event, subsequent overtopping can occur as the flow oscillates in the basin, a shock being reflected between the back-wall and the barrier. These oscillations may cause the depth at the barrier to exceed $B$ again at later times, indeed in \cref{fig:ot:CritWall_Fin_Profiles}(g,h) both simulations exhibit a secondary overtopping when the shock collides with the barrier. It is possible for the oscillations to result in many overtopping events. 

During this oscillating phase of motion, we may be concerned that mixing becomes important. The mixing at the upper shear layer can be shown to be small by an analysis of mixing rate reported in \citet{ar_Strang_2001}. Across the shock, though, the depth approximately doubles during the first passage of the shock across the domain (\cref{fig:ot:CritWall_Fin_Profiles}(a,b)), becoming substantially weaker after the first reflection due to energy dissipation. As shown by \citet{ar_Borden_2012} for moving shocks and \citet{ar_Wood_1984} and \citet{ar_Lawrence_2022} for stationary shocks, shocks of this size do not cause substantial mixing, and instead the principle difference between our model and real currents is in the balance of momentum flux in the single layer model \cref{eqn:ot:SW_shock_con_general_momentum} due to the energy dissipation and inertia in the upper layer. We do not expect this effect to be significant in our case of interest, flows under a deep ambient, but certainly would need to be included if the ambient and current were of a similar depth.

\begin{figure}
	\centering
	\includegraphics{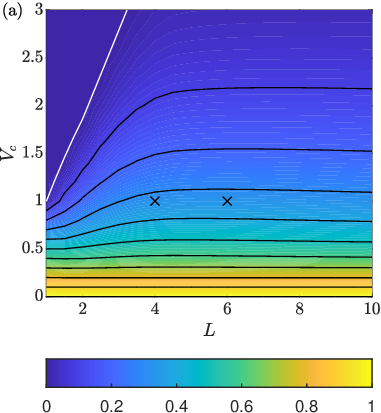}
	\includegraphics{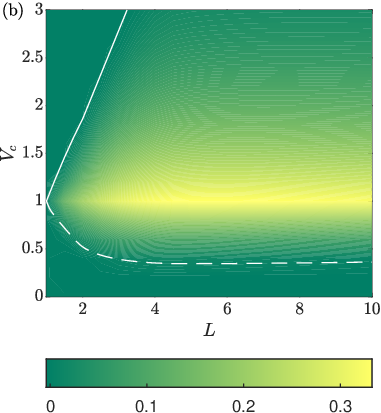}
	\caption{(a) The escaped volume $\Delta V_\infty$ (the colour-bar) as a function of $L$ and $V_c$ for $\Fro = \sqrt{2}$. The black lines are the contours $\Delta V_\infty \in \{ 0.1,0.2,\ldots,1 \}$. We also plot in a white solid curve the maximum $V_c$ for overtopping from \cref{fig:ot:InfWall_MaxDepth}. The crosses mark the parameter values used in \cref{fig:ot:CritWall_Fin_Profiles}. (b) The difference between the trapped volume and the value predicted by a purely geometric argument, $\min(V_c,1) - V_\infty$. We plot in a white solid curve the maximum $V_c$ for overtopping from \cref{fig:ot:InfWall_MaxDepth}, and in a white dashed curve the function \cref{eqn:ot:CritOvertopping_Inertia_BL} with $R=8$ which approximately bounds the region in which $V_\infty = V_c$.}
	\label{fig:ot:OutflowVolume_Crit}
\end{figure}

To characterise the overtopping, we calculate the dimensionless volume of fluid that has escaped at late times
\begin{align}\label{eqn:ot:escaped_volume}
	\Delta V &= 1 - V_\infty,
	&&\text{where}&
	V_\infty &= \lim_{t \rightarrow \infty} \int_{0}^L h \dd{x}.
\end{align}
Whilst formally $V_\infty$ should be evaluated at $t \rightarrow \infty$, we compute it numerically by stopping the simulation when there is less than $0.001$ change in volume over $2$ oscillations of the fluid trapped behind the barrier. Specifically, at the instant of collision the volume of fluid in the domain is calculated as $V_0 = 1$, then the simulation is run for a period $L^{3/2}/V_0^{1/2}$, which approximates the time for a characteristic to cross the domain. This is then repeated, $V_n$ being the volume after $n$ runs, and the duration of the $(n+1)^\nth$ run being $L^{3/2}/V_n^{1/2}$. If, after $V_n$ is calculated, we have $\max_{m \in \{n-4 , \ldots , n\}}(V_m) - \min_{m \in \{n-4 , \ldots , n\}}(V_m) < 0.001$, then we cease performing further computations, and approximate $V_\infty \approxeq V_n$. We simulate for $L \in \{1,1.5, \ldots ,10\}$, $V_c \equiv BL\in \{0.2,0.4, \ldots ,3\}$ with a resolution of 1000 cells, and include in our plots that all the fluid escapes for $B=0$, and that the fluid drains to a height $h=B$ for $L=1$ as discussed in \citet{ar_Skevington_F001_Draining}. We now explore the overtopping that occurs for $L>1$, for which the escaped volume is shown in \cref{fig:ot:OutflowVolume_Crit}(a).

The simplest limiting case is that $\Delta V = 0$ when no fluid is able to overtop, thus $V_\infty = 1$. For smaller barriers which permit overtopping, it is possible for the limiting volume to be exactly that required by volume conservation, $V_\infty = V_c \leq 1$, thus $\Delta V = 1 - V_c$. For this to happen the limiting behaviour must be that discussed in \citet{ar_Skevington_F001_Draining}, \ie at late times
\begin{align}
	h &\sim h_a \eqdef B + \frac{27}{2 L^2 \tau^2} &
	uh &\sim q_a \eqdef \frac{27 x}{L^4 \tau^3}
\end{align}
where $\tau$ is an offset time. For the proper ordering of terms the difference between the leading order solution above and the exact solution cannot be too large. We suppose that, at the instant the current reaches the barrier, $h_a L = 1$ which determines $\tau$, thus
\begin{align}
	q_a = \ppar*{ \frac{2 (1 - V_c)}{3} }^{3/2} \frac{x}{L^{5/2}}.
\end{align}
We then compare the total inertia of the flow at the instant of collision $Q$ (from the simulation of a lock-release current) to that of the asymptotic $Q_a$, producing the ratio
\begin{flalign}
&&	&&
	R &\eqdef \frac{Q}{Q_a},	\\
&\text{where}&
	Q &\eqdef \int_0^L uh \dd{x},	
	&
	Q_a & \eqdef \int_0^L q_a \dd{x} = \ppar*{\frac{2}{L}}^{1/2} \ppar*{ \frac{(1 - V_c)}{3} }^{3/2},	& \\
&\text{thus}&	&&
	V_c &= 1 - \frac{3 L^{1/3} Q^{2/3}}{2^{1/3} R^{2/3}}.
	\label{eqn:ot:CritOvertopping_Inertia_BL}
\end{flalign}
Our simulations suggest that, so long as $R \lesssim 8$ then $V_\infty \approx V_c$, see \cref{fig:ot:OutflowVolume_Crit}(b). We see then that the requirement for limiting to the draining discussed in \citet{ar_Skevington_F001_Draining} is that the inertia is not too much greater than that of the asymptotic solution, $Q \lesssim 8 Q_a$. This bound is surprisingly weak, a priori it may be anticipated that $\abs{R - 1} \ll 1$ is required; it seems the asymptotic solution is unusually robust.

This leaves an intermediate regime, where the inertia is large enough to cause overtopping in excess of that required by volumetric arguments, and the barrier insufficiently tall to prevent outflow. In this regime the inertia is enough to decrease the remaining volume below that which is required by simple volumetric arguments, $V_\infty = \min(V_c,1)$. Indeed, for $V_c \approx 1$, $L \gtrsim 3$ the outflow volume is over $0.3$ greater than the minimal value, which shows the inertia has a significant effect. In this regime the fluid oscillates in the basin, and during each oscillation some portion of the fluid overtops (\cref{fig:ot:CritWall_Fin_Profiles}).

Beyond $L=10$, indeed largely beyond $L=5$, the current is approximately in similarity form at collision and, to leading order, $\Delta V$ is a function of $V_c$. This is because the characteristic reflected off the back-wall has caught up with the front, see \cref{fig:ot:CritWall_Fin_Profiles}. This means that, so long as $V_c$ remains constant, the barrier may be placed any distance downstream and the same behaviour will be observed.

Our analysis shows that a barrier constructed so that $V_c = 1$ is not capable of containing the fluid. That is, if the confined volume is equal to the volume of the current then we should expect about 30\% of the fluid to escape. From our simulations, we expect that a barrier confining around 2 to 3 times the volume of the current is required to stop it, and even then a small portion of the current, around 5 to 10\%, will still overtop.

\subsection{Dam-break flow: $\Fro \rightarrow \infty$} \label{sec:overtop_simulation_dambreak}

\begin{figure}
	\centering
	\includegraphics{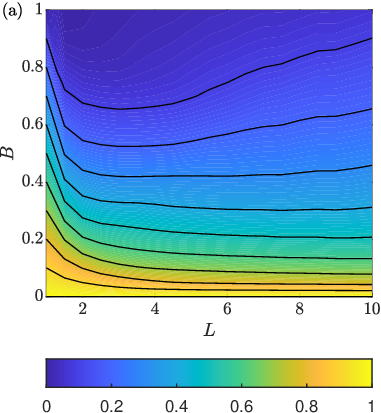}
	\includegraphics{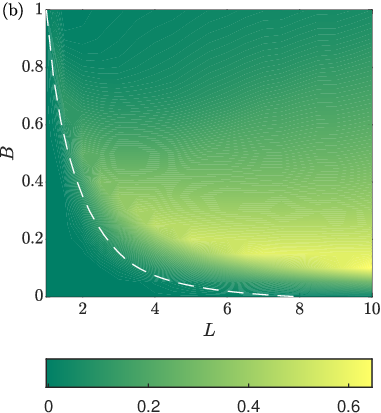}
	\caption{(a) The escaped volume $\Delta V_\infty$ (the colour-bar) as a function of $L$ and $V_c$ for $\Fro \to \infty$. The black lines are the contours $\Delta V_\infty \in \{ 0.1,0.2,\ldots,1 \}$. (b) The difference between the trapped volume and the value predicted by a purely geometric argument, $\min(V_c,1) - V_\infty$. We plot in a white dashed curve the function \cref{eqn:ot:CritOvertopping_Inertia_BL} with $R=8$ which approximately bounds the region in which $V_\infty = V_c$.}
	\label{fig:ot:OutflowVolume_CritInf}
\end{figure}

In this case the current does not evolve into similarity form, and the front is not caught up by the reflected characteristic from the back-wall \citep{ar_Hogg_2006}. Thus the $\beta$-fan \cref{eqn:ot:lockrel_early_soln} generated by the initial release persists for all time. Hence, the initial overtopping behaviour is independent of $L$ and only depends on $B$ as shown in \cref{fig:ot:collision_close} for $\Fro \to \infty$. Specifically, supercritical overtopping occurs for $B < 2$, though the volume that escapes may be quite small prior to the transition to other modes.

The escaped volume as measured using \cref{eqn:ot:escaped_volume} is plotted in \cref{fig:ot:OutflowVolume_CritInf}(a) from numerical simulations. We observe that, for $L \gtrsim 5$ and $B \lesssim 0.5$, $\Delta V_\infty$ is only weakly dependent on $L$. This is because the reflected characteristic from the back is advancing slowly across the $\beta$-fan. For $L < 5$ the effect of the back-wall is more pronounced, thus the dependence on $L$ is stronger. Similarly to the Boussinesq case, the boundary of the region in which $V_\infty \approx V_c$ is given by \cref{eqn:ot:CritOvertopping_Inertia_BL} with $R=8$ ($Q$ evaluated for a dam-break current), see \cref{fig:ot:OutflowVolume_CritInf}(b). However, in this case the region is much smaller. Indeed, across a wide range of parameters the actual escaped volume is substantially underestimated by volumetric considerations; in excess of an additional $60\%$ of the initial volume overtops the barrier for $L=10$, $B=1/10$.

%% file: Sections/Experiment.tex
\section{Comparison to experiments and 3D simulations} \label{sec:overtop_experiment}

\subsection{Dam break flow}

The boundary condition used in this study is very similar to that developed by \citet{ar_Cozzolino_2014}, only differing in that they chose to suppress supercritical overtopping. They compared the predictions of their model to a dam-break ($\Fro\to\infty$) experiment performed by \citet{ipr_Hiver_2000} with $L=1.74$, $B=0.53$, $\epsilon=0.19$, $\theta=7.6\degree$, $H/X=0.048$, and found a good agreement. However, we note that the overtopping predicted for this set of parameter values features only a very short period of supercritical overtopping, and so does not clearly differentiate between the two boundary conditions.

Instead, we compare our model to the experiments of \citet{ar_Greenspan_1978}. In particular, their data evidences an initial period of supercritical overtopping, as is modelled by our boundary condition. Their figure 9 shows a dam-break experiment with $L=2$, $B=0.5$, $\epsilon=0.26$, $\theta=60\degree$, $H/X=0.89$, a jet can be seen that initially leaps the barrier prior to the formation of an upstream bore. 

\begin{figure}
	\centering
	\includegraphics{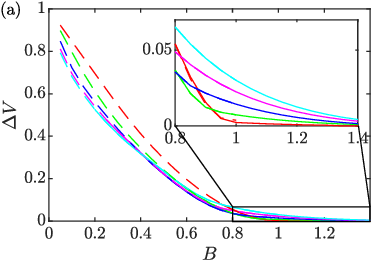}
	\includegraphics{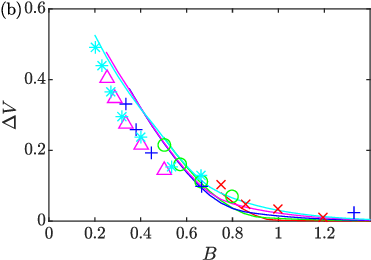}
	\includegraphics{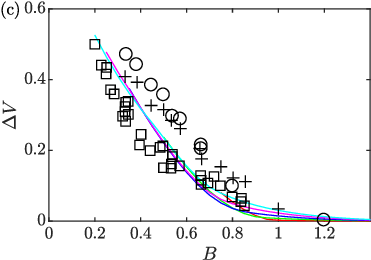}
	\includegraphics{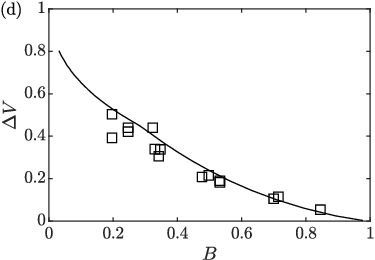}
	\caption{Comparison between our theoretical predictions (\cref{sec:overtop_simulation_dambreak}) and the experimental results of \citet{ar_Greenspan_1978}. In all plots the escaped volume is plotted as a function of barrier height; the curves are our simulations, and the markers are experiments. 
	In (a) and (b) the different distances to the barrier $L$ correspond to the plot colours and markers: red $\times$, $4/3$; green $\bigcirc$, $2$; blue $+$, $3$; magenta $\triangle$, 4; cyan $\ast$, 5. In (a) we plot only numerical results, the solid curves being the volume that overtops during the initial period of overtopping, the dashed representing the late time cumulative escaped volume. In (b) we plot the same solid curves, along with experimental results for a barrier at $\theta=90\degree$.
	In (c) we plot the same solid curves, but now experiments are shown for different angles $\theta$: $\bigcirc$, $30\degree$; $+$, $60\degree$; $\square$, $90\degree$. 
	In (d) the experiments are for $\theta=90\degree$, but in this panel $L=1/B$.}
	\label{fig:ot:Greenspan}
\end{figure}

The dam-break experiments of \cite{ar_Greenspan_1978} only explored the initial period of overtopping for the case $V_c \geq 1$. In this parameter regime the fluid is below the elevation of the barrier when the surface is level, and consequently overtopping is driven by the oscillatory sloshing motion depicted in \cref{fig:ot:CritWall_Fin_Profiles}. The initial period of overtopping is the time up to the instant when the depth drops from $h>B$ to $h \leq B$ at $x=L^-$ (if $h<B$ for all $t$ then then initial period extends to $t\to\infty$). Both the volume that overtops during the initial period and the volume that has escaped by late times are plotted in \cref{fig:ot:Greenspan}(a) from simulations of our model, showing that there is little difference between these two measures (at least, for the range of $B$ and $L$ plotted). Additionally, as noted by \citet{ar_Greenspan_1978}, the curves for different $L$ are very similar. We understand this as the effect of the $\beta$-fan in the frontal region of the dam-break that persists over long distances \citep{ar_Hogg_2006}. The extent of this frontal region relative to the length of the overall current varies slowly as it is caught up by the reflection of the $\beta$-fan from the back-wall.

We next consider the experiments with a barrier at $\theta = 90\degree$. There may be concerns that the deviation of the critical barrier boundary condition precludes application to such steep slopes, however we only require that there is no energy difference between the `base' and `crest'. These are not necessarily at the locations immediately adjacent to the incline, but rather the first locations up- and down-stream where the assumptions of the shallow water equations are satisfied, thus the model may be applied with a small change in interpretation. The experiments show good agreement with our simulations (\cref{fig:ot:Greenspan}(b)), with only small discrepancies. For the barriers that are close and tall (red) the experiments show a slightly larger escaped volume than predicted by our simulations, and we suspect that this discrepancy is because the region around the barrier where the shallow water assumptions are violated is large in comparison to the domain ($\epsilon \not\ll 1$). For the shorter, distal barriers (cyan and magenta) our simulations over-predict the escaped volume, and it is likely that this is in part because of drag reducing the inertia of the physical current, which will reduce the escaped volume. Indeed, comparing our simulations to the experiments of \citet{ipr_Hiver_2000} we find that our initial overtopping exceeds that found experimentally, while the example simulation with drag reported by \citet{ar_Cozzolino_2014} showed better agreement. In the regime $V_c=1$ ($L=1/B$), however, our simulations agree remarkably well with experiment (\cref{fig:ot:Greenspan}(d)), indicating that the small errors cancel in this important case.

Investigating the effect of different barrier angles $\theta$, we find that our simulations slightly underestimate the escaped volume for $\theta=30\degree$ and $\theta=60\degree$. These experiments were not performed at the scales at which our model was derived (see \cref{sec:overtop_PF}), in particular $\epsilon \not\ll 1$ and $Z \not\ll X$, thus higher order corrections to the model may be required to fully capture these non-shallow dynamics.

\subsection{Boussinesq gravity current} \label{sec:overtop_experiment_bouss}

\begin{figure}
	\centering
	\includegraphics{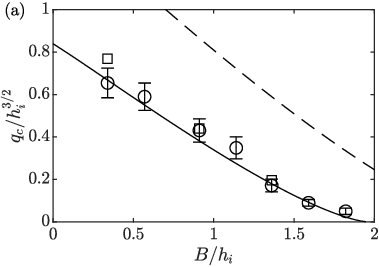}
	\includegraphics{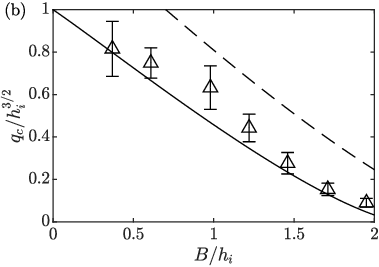}
	\caption{The overtopping flux, $q_c/h_i^{3/2}$, as a function of barrier height, $B/h_i$, both scaled relative to the depth of the incident fluid, $h_i$. Potted are data from the simulations of \citet{ar_GonzalezJuez_2009b} ($\square$, $\Rey=3535$; $\bigcirc$/$\triangle$ with error bar, $\Rey=707$) and the predictions of the analysis in \cref{sec:overtop_close} (solid lines). In (a) the simulations have no-slip boundary conditions and $\Fro=0.83$, while (b) have slip boundaries and $\Fro=1.00$. Also plotted in both are the predictions with $\Fro = \sqrt{2}$ (dashed lines).}
	\label{fig:ot:GonzalezJuez}
\end{figure}

Our analysis applies to overtopping currents, propagating under relatively deep ambient fluid so that the momentum of the latter may be neglected.  This condition is particularly 
important when interpreting data from laboratory and numerical simulations of Boussinesq currents, many of which feature currents that are not initially shallow relative to the ambient and are potentially strongly influenced by the motion of the ambient fluid.  Notably, however, \citet{ar_GonzalezJuez_2009b} report two-dimensional, direct numerical simulations of gravity currents driven from a sustained source and propagating under a deep ambient towards an obstacle that spans the channel.  
They present data on the depth and velocity of the gravity current motion downstream of the obstacle (their figure 12), from which we can compute the overtopping flux of fluid.  We present their simulation results in \cref{fig:ot:GonzalezJuez}, along with the outflow flux predicted by our solution of the boundary-Riemann problem in \cref{sec:overtop_close}, in which we use the Froude number of the incident fluid, $F_i\equiv u_i/h_i^{1/2}$,  determined by \citet{ar_GonzalezJuez_2009b}.  
The data series with no-slip and slip basal boundary conditions have different reported frontal Froude numbers.  In both cases, we observe in \cref{fig:ot:GonzalezJuez} that there is quite close agreement between the simulated results and the predictions from our boundary-Riemann calculation.  We also note that the analysis in \cref{sec:overtop_close} gives a formal justification for the simplified modelling used by \citet{ar_GonzalezJuez_2009b}; their box model analysis upstream  of the barrier is identical to the full solution of a shallow  water model with the critical barrier boundary condition derived in \cref{sec:overtop_PF}, and their simulations all fall into regime A-II of the boundary-Riemann problem of \cref{sec:overtop_close} (see \cref{fig:ot:CritWall_BndRie}).

More complete model comparisons with data from simulations and experiments would enable us to test other aspects of our predictions, such as flow depths and velocities and shock speeds and to probe dynamical effects that we have not included.
The three-dimensional, large eddy simulations of gravity currents flowing over a triangular obstacle performed by \citet{ar_Tokyay_2015} report some of these features.
However, we find that the depth computed above the crest of the obstacle differs substantially from the critical depth, as calculated using their reported data. 
It is possible  that this discrepancy comes from how the depth and velocity of the current have been calculated: the former is actually the depth integral of concentration (normalised by the concentration at release), and the latter the concentration flux per unit `depth'.
Consequently, any mixing present prior to overtopping has
a substantial effect on the measurement, and the initial conditions used to generate  these
currents are likely to cause a substantial mixed layer to form.
Some recent laboratory experiments have been performed by \citet{ar_DeFalco_2021} and \citet{ar_Adduce_2022} using lock release of saline water propagating through fresh water and overtopping an initially distant triangular obstacle.  The width-averaged density of the the flowing current is measured, as well as the position of the front and the proportion of the released fluid that is retained behind the barrier.  The currents are released from a lock that occupies the full depth of the ambient; therefore the motion is strongly influenced by flow within the ambient.  Evidence of this is provided from the measurement of the speed of the front, which is approximately constant and close to the predictions of two-layer hydraulic models (see Appendix B of \citet{ar_Hogg_2016} for the analytical evaluation of the front speed in a two-layer model of slumping).  The effects of the upper layer are not included in the model of overtopping developed here and this precludes comparison with this dataset. 

%% file: Sections/Conclusion.tex
\section{Summary and conclusions} \label{sec:conclusion}

The collision of a gravity current with a barrier has been investigated using the shallow water equations, revealing the spatial and temporal dependence of the fluid confined behind the barrier, and the net volume of fluid that is transported over it following release from a lock.

These calculations have utilised a sophisticated boundary condition to model the effects of a barrier, modifying that of \citet{ar_Cozzolino_2014} and generalising that of \citet{ar_Skevington_F001_Draining}. The resulting model has been explored in the case of a collision with a spatially and temporally uniform current, \cref{sec:overtop_close}, which is a similar flow scenario to the studies of \citet{ar_Long_1954,ar_Long_1970} and \citet{bk_Baines_TESF}, except that we require that the flow beyond the barrier is supercritical which yields a unique solution. In particular, we find for any uniform state adjacent to the barrier, one of six behaviours will be exhibited (\cref{fig:ot:CritWall_BndRie}): uniform supercritical overtopping (A-I); an upstream propagating bore with subcritical overtopping (A-II); an upstream propagating bore with blocked flow (A-III); a rarefaction fan with subcritical overtopping (A-IV); a rarefaction fan with blocked flow (A-V); or a rarefaction fan leaving the barrier dry (A-VI). The classification of the interaction has been extended to the case of an unsteady and spatially varying current. In particular, we analyse the transitions between them for a lock-release current. This has been carried out for barriers that are relatively close to the release (\cref{sec:overtop_collision_close}) and relatively distant (\cref{sec:overtop_collision_far}), revealing the consequences of the dynamics in the oncoming current on the overtopping.

In addition to the analytical results we presented numerical simulations in \cref{sec:overtop_simulation}. These demonstrated that, for a `lock-release' flow with a specified frontal Froude number, the volume of fluid that traverses the barrier depends on the dimensionless distance from the back-wall to the barrier, $L$, and the dimensionless barrier height, $B$. It was shown that, for Boussinesq currents where the reflected characteristic off the back-wall quickly catches up with the front, the escaped volume is approximately a function of only $V_c \equiv BL$ for $L \geq 5$, thus for these distances the current may already be considered to approximately be in similarity form for the purposes of the collision (\cref{sec:overtop_collision_far}). Conversely, for a dam-break current, the escaped volume is approximately a function of only $B$ for $L \geq 5$, $B \leq 1/2$, showing that the effects of the back-wall do not significantly affect the overtopping dynamics in this regime.

An important application of our results is in the design of barriers to trap fluids in the case of spillage or rupture of a container. We find that, for a Boussinesq current, a barrier that confines precisely the volume of fluid may permit $30\%$ of the fluid to escape. Instead, a barrier that could confine a substantial volume of quiescent fluid, perhaps 2 to 3 times the volume of released fluid, is required to contain the current. Even with such a barrier, a small portion of the fluid will still overtop, meaning that additional defences are required downstream to totally stop the current. However, in real flows mixing may be important, and in situations where the upper ambient fluid is also shallow a principle difference is expected to be the influence of the inertial and energetic processes in the upper layer. Further research is required to quantify these effects. For a dam-break current, a barrier that confines precisely the volume of the released fluid may have $60\%$ or more of the fluid overtop, and to contain the current a barrier of similar height to the initial release is required. In this case, we expect the effect of drag to have a non-negligible effect over the distances considered, and again further investigation is required to quantify its effect.

Our comparison to experimental and simulation data in \cref{sec:overtop_experiment} provided support for our theoretical model, but highlighted some additional features that could be included in future models. Indeed, high quality datasets would be of very considerable value for future testing and refinement of the model. Firstly, it is believed that drag will slightly reduce the magnitude of overtopping in the range of parameters considered, and it is possible that in other regions of parameter space the magnitude of the effect could be larger; it remains to classify the effect of drag on the overtopping of both dam-break and lock-release flows. Secondly, we argued that the likely cause of the discrepancies in our comparison to experiment was the violation of the assumptions about length-scales made in the original derivation; how best to include higher order corrections in the model? Finally, for the case of a gravity current there is the possibility that entrainment of ambient, and the inertia of the upper layer, may have an influence on and be influenced by the overtopping process. The fundamental theory presented here for the canonical, simplified problem forms a basis for these future research endeavours.

%% file: Sections/Appendix_Const_Surface.tex
\section{The constant surface elevation boundary condition} \label{sec:overtop_app_constsurface}

In some earlier studies \citep[\eg][]{ar_Greenspan_1978,ar_Rottman_1985}, subcritical flow over a barrier is modelled on the assumption that the surface elevation is constant between the base ($x=L_1$) and the crest ($x=L_2$), along with the condition of criticality ($h = q^{2/3}$) at the crest. In these models
\begin{equation}
	\eval*{h}_{x=L_1} = \eval*{h}_{x=L_2} + B = \eval*{q^{2/3}}_{x=L_2} + B = \eval*{(uh)^{2/3}}_{x=L_1} + B,
\end{equation}
thus the boundary condition is
\begin{equation}
	h_b-B = (u_b h_b)^{2/3}.
\end{equation}
While this may appear reasonable, in fact it predicts energy generation. Indeed by \cref{eqn:ot:BC_energy_loss} the energy loss is required to be positive, $\Delta E_b \geq 0$, whereas the above model predicts a negative value
\begin{align}
	\Delta E_b 
	&= \textfrac{1}{2} u_b^2 + h_b - B - \textfrac{3}{2} (u_b h_b)^{\frac*{2}{3}}
	= \textfrac{1}{2} u_b^2 - \textfrac{1}{2} (u_b h_b)^{\frac*{2}{3}}
	\notag\\
	&= \textfrac{1}{2} (u_b h_b)^{\frac*{2}{3}} \ppar*{ F_b^{4/3} - 1 } < 0.
\end{align}
Here $F_b = u_b/h_b^{1/2}$ is the local Froude number at the base of the slope, and for subcritical conditions at that location $0 < F_b < 1$.

We may ask: what boundary condition should we impose if we wish to minimise the change in surface elevation while not violating conservation of energy? That is we choose to impose 
\begin{equation}
	\Delta E_b = \frac{u_b^2}{2} + h_b - B - \frac{3}{2} (u_b h_b)^{\frac*{2}{3}}
\end{equation}
where $\Delta E_b \geq 0$ is some value to be determined so as to minimise
\begin{equation}
	\Delta h \eqdef \eval*{h}_{x=L} - (\eval*{h}_{x=L+\epsilon} + B) = h_b - B - (u_b h_b)^{2/3}.
\end{equation}
This is the type of problem considered in \cref{sec:overtop_app_Riemann}, for a given set of initial conditions in a boundary-Riemann problem how best to choose $\Delta E_b$ to minimise $\Delta h$? By \cref{eqn:ot:diff_energy_wrt_height_drop} we have that $\pdv*{\Delta h}{\Delta E_b}>0$ (holding initial condition constant, only changing boundary values), thus the minimal drop in elevation permitted by energetic considerations is that found by imposing $\Delta E_b=0$ for subcritical outflow, as is done in this study (see \cref{sec:overtop_PF}).

%% file: Sections/Appendix_BoundRiemann_CritWall.tex
\section{Results pertaining to the boundary-Riemann problem}	\label{sec:overtop_app_Riemann}

\subsection{Classification of solutions, with existence and uniqueness}	\label{sec:overtop_classify}

\begin{figure}
	\centering
	\includegraphics{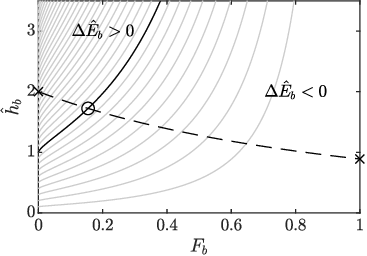}
	\caption{An illustration of the condition for subcritical overtopping to exist. The solid black curve represents the condition $\Delta \hat{E}_b = 0$, and divides the plane into two regions in which $\Delta \hat{E}_b$ is of differing sign. The solid grey curves are contours $\Delta \hat{E}_b \in \{ -0.9,-0.8,\ldots,3.4 \}$. The dashed curve represents the flow states at $\hat{x}=0$ accessible from the initial condition $\hat{u}_i = 0$, $\hat{h}_i = 2$. The change in sign on the dashed curve between $\hat{u}_b = 0$ and $\hat{u}_b = \hat{h}_b^{1/2}$, marked with crosses, implies the existence of an accessible flow state satisfying $\Delta \hat{E}_b = 0$, marked with a circle, which is the flow state that solves the boundary-Riemann problem.}
	\label{fig:ot:CritWall_Energy}
\end{figure}

Here we discuss the classification of solutions which do not correspond to supercritical overtopping, thus $\hat{u}_i \leq \hat{h}_i^{1/2}$ or $\Delta \hat{E}_i \leq 0$ (or both), and do not contain a dry region adjacent to the boundary, thus $\hat{u}_i > -2 \hat{h}_i^{1/2}$. We explore existence and uniqueness of solutions. A priori, it is not clear whether a given initial condition $(\hat{u}_i,\hat{h}_i)$ meeting the preceding constraints will have a solution subject to applying the subcritical overtopping condition from \cref{sec:overtop_PF}, and we may be concerned by the possibility that multiple solutions exist. In this appendix we find when a solution to subcritical overtopping exists, and demonstrate that it is unique. 

For the purpose of this discussion, we define a \emph{flow state} to be any solution to the shallow water system \cref{eqn:ot:SW_sys} satisfying the initial conditions \cref{eqn:ot:overtop_close_ic} irrespective of any boundary conditions. Thus there are infinitely many flow states, each being either a uniform state \cref{eqn:ot:boundRie_soln_steady}, a $\beta$-fan \cref{eqn:ot:boundRie_soln_fan}, or a $\beta$-shock \cref{eqn:ot:boundRie_soln_shock}. For a given initial condition $(\hat{u}_i, \hat{h}_i)$ the flow states form a one parameter family, and during our discussion we will employ the following parameters one by one: $F_b \eqdef \hat{u}_b / \hat{h}_b^{1/2}$, $\hat{\mu}_b \eqdef \hat{u}_b - \hat{h}_b^{1/2}$, and $\hat{s} \eqdef (\hat{u}_i \hat{h}_i - \hat{u}_b \hat{h}_b) / (\hat{h}_i - \hat{h}_b)$. Each parameter uniquely specifies the solution by the results in \cref{sec:overtop_dE}, so long as they are in a regime so that dynamics at the outflow are at most critical, thus $F_b \leq 1$, $\hat{\mu}_b \leq 0$. For $\hat{s}$, we first note that it is still a defined quantity for the case of a $\beta$-fan, though it no longer represents a shock speed. To specify a solution uniquely, we require
\begin{subequations}
\begin{flalign}		
&&
	\label{eqn:ot:app_shock_max}
	\hat{s} &\leq \frac{ \hat{u}_{bc}^3 - \hat{u}_i \hat{h}_i}{\hat{u}_{bc}^2 - \hat{h}_i}
\\&\text{where}&
	\label{eqn:ot:app_critvel_fan}
	\hat{u}_{bc} &= \frac{\hat{u}_i + 2 \hat{h}_i^{1/2}}{3}
	&&\text{for}&
	\hat{u}_i &\leq \hat{h}_i^{1/2},
&\\&\text{and}&
	\label{eqn:ot:app_critvel_shock}
	\hat{u}_{bc} - \hat{u}_i &= - \ppar*{\frac{\hat{u}_{bc}^2 + \hat{h}_i}{2 \hat{u}_{bc}^2 \hat{h}_i}}^{1/2} (\hat{u}_{bc}^2 - \hat{h}_i)
	&&\text{for}&
	\hat{u}_i &\geq \hat{h}_i^{1/2},
\end{flalign}
\end{subequations}
specify the velocity at the barrier under critical flow conditions $\hat{u}_{bc}$, the value of $\hat{u}_b$ when $\hat{u}_b=\hat{h}_b^{1/2}$: \cref{eqn:ot:app_critvel_fan} is a consequence of $\alpha$ being constant across a $\beta$-fan; \cref{eqn:ot:app_critvel_shock} is from the shock conditions \cref{eqn:ot:SW_shock_con_general}; and \cref{eqn:ot:app_shock_max} is because of \cref{eqn:ot:app_diff_shockspeed_by_vel}.

First, we consider flow states parametrised by $F_b$, thus $\Delta \hat{E}_b$ is a function of $\hat{u}_i$, $\hat{h}_i$, $F_b$. The question as to whether there exists a flow state satisfying $\Delta \hat{E}_b = 0$ can be answered by showing that there exists an interval in $F_b$ over which $\Delta \hat{E}_b$ changes sign. Restricting ourselves to subcritical flow $0 \leq F_b \leq 1$, for $F_b = 0$ we find $\Delta \hat{E}_b = \hat{h}_b - 1$ whilst $F_b = 1$ yields $\Delta \hat{E}_b = - 1$. Thus, if the flow state with $F_b = 0$ satisfies $\hat{h}_b \geq 1$ then a solution exists to imposing $\Delta \hat{E}_b = 0$ (intermediate value theorem), this is illustrated in \cref{fig:ot:CritWall_Energy}. We can further show the existence of a flow state with $F_b = 0$, $\hat{h}_b \geq 1$ is necessary for $\Delta \hat{E}_b=0$ to have a solution by noting that $\pdv*{\Delta \hat{E}_b}{F_b} < 0$ for $0 < F_b < 1$, see \cref{sec:overtop_dE}, which also shows the solution is unique.

Next we seek to divide up the $(\hat{u}_i,\hat{h}_i)$-plane based on the type of solution found in each region. If blocked flow is imposed then $\hat{u}_i = 0$ results in a uniform state, $\hat{u}_i < 0$ a $\beta$-fan, and $\hat{u}_i > 0$ a $\beta$-shock. For subcritical outflow the situation is more complicated. If $\hat{u}_i<0$ then $\hat{u}_i \leq \hat{u}_b$ thus we have a $\beta$-fan, and if $\hat{u}_i \geq \hat{h}_i^{1/2}$ then $\hat{u}_i > \hat{u}_b$ thus we have a $\beta$-shock. For subcritical initial conditions ($0 < \hat{u}_i < \hat{h}_i^{1/2}$) we consider $\Delta \hat{E}_b$ a function of $\hat{u}_i$, $\hat{h}_i$, $\hat{\mu}_b$, from which we find that $\pdv*{\Delta \hat{E}_b}{\hat{\mu}_b} < 0$ for $0 < \hat{u}_b < \hat{h}_b^{1/2}$, see \cref{sec:overtop_dE}. We take the uniform flow state and move through parameter space to the $\Delta \hat{E}_b = 0$ flow state, thus if $\Delta \hat{E}_i > 0$ then $\hat{\mu}_i < \hat{\mu}_b$ and we have a $\beta$-fan, whilst if $\Delta \hat{E}_i < 0$ then $\hat{\mu}_i > \hat{\mu}_b$ and we have a $\beta$-shock.

The final thing to check is that the dynamics derived above all occur within the domain. That is to say, if in the above construction we inadvertently made use of a shock or fan that existed in the region $\hat{x}>0$ then the results would be invalid. To demonstrate that the dynamics are confined to $\hat{x} \leq 0$ we note that in a uniform state there is no spatial variation, and for solutions with a $\beta$-fan the fact that we enforce subcritical or blocked flow implies that $\hat{\mu}_b < 0$, thus a uniform region exists between the fan and the barrier. Therefore the only cases that require any computation are those with $\beta$-shocks. The shock speed is given by \cref{eqn:ot:boundRie_soln_shock}, and because $\hat{h}_b > \hat{h}_i$ we know that the shock satisfies $\hat{s}<0$ precisely when $\hat{u}_b \hat{h}_b < \hat{u}_i \hat{h}_i$, which is clearly satisfied for the blocked case. For subcritical overtopping, when the initial condition is subcritical $\hat{u}_i \leq \hat{h}_i^{1/2}$ then, because the $\beta$-shock absorbs $\beta$-characteristics (Lax entropy), we have $\hat{s}<0$. When $\hat{u}_i > \hat{h}_i^{1/2}$ then we know $\Delta \hat{E}_i \leq 0$ or else the solution would be supercritical overtopping. Thus we can construct a flow state with $\hat{s}=0$ for which $\Delta \hat{E}_b < \Delta \hat{E}_i \leq 0$, where we have used that $\shock{E}<0$ for a stationary $\beta$-shock. The flow downstream of this stationary shock shock is subcritical. Treating $\Delta \hat{E}_b$ a function of $\hat{u}_i$, $\hat{h}_i$, $\hat{s}$, for $0 < \hat{u} < \hat{h}^{1/2}$ we can show $\pdv*{\Delta \hat{E}_b}{\hat{s}}<0$, see \cref{sec:overtop_dE}, therefore to obtain the state $\Delta \hat{E}_b = 0$ we must decrease $\hat{s}$ and thus, in the required solution, $\hat{s}<0$.

\subsection{Derivatives of the energy difference at the barrier}	\label{sec:overtop_dE}

We consider an arbitrary flow state for a given, fixed initial condition $\hat{u}_i$ $\hat{h}_i$, the flow states parametrised by the velocity at the barrier $\hat{u}_b$. This means that we can consider $\hat{h}_b$ and $\Delta \hat{E}_b$ as functions of $\hat{u}_i$, $\hat{h}_i$, $\hat{u}_b$. To start with we evaluate the derivative of $\Delta \hat{E}_b$ with respect to $\hat{u}_b$ from which we can determine its sign. By \cref{eqn:ot:BC_energy_loss}
\begin{align}	\label{eqn:ot:app_dEdu_chain}
	\pdv{\Delta \hat{E}_b}{\hat{u}_b} &= \ppar*{ F_b - \frac{1}{F_b^{1/3}} } \hat{h}_b^{1/2} + ( 1 - F_b^{2/3} ) \pdv{\hat{h}_b}{\hat{u}_b},
\end{align}
where $F_b \eqdef \hat{u}_b / \hat{h}_b^{1/2}.$ From here we require $\pdv*{\hat{h}_b}{\hat{u}_b}$. The two types of solution will be considered separately. For a $\beta$-fan we use that $\alpha$ is constant across the domain, that is
\begin{subequations}
	\begin{align}
	\hat{u}_b + 2 \hat{h}_b^{1/2} &= \hat{u}_i + 2 \hat{h}_i^{1/2},
	&&\text{and so}&
	\pdv{\hat{h}_b}{\hat{u}_b} &= - \hat{h}_b^{1/2}.
\end{align}
For a $\beta$-shock we can deduce from \cref{eqn:ot:SW_shock_con_general} that
\begin{align}
	\hat{u}_b - \hat{u}_i &= - \ppar*{ \frac{\hat{h}_b + \hat{h}_i}{2 \hat{h}_b \hat{h}_i}}^{1/2} (\hat{h}_b - \hat{h}_i),
	&&\text{and so}&
	\pdv{\hat{h}_b}{\hat{u}_b} &= - \frac{\ppar*{8 H (H+1)}^{1/2}}{H^2 + H + 2} \hat{h}_b^{1/2}.
\end{align}
where $H \eqdef \hat{h}_i/\hat{h}_b$, and $0 < H \leq 1$ for a $\beta$-shock. We combine these two cases into the single expression
\begin{align}\label{eqn:ot:app_dhdu}
	\pdv{\hat{h}_b}{\hat{u}_b} &= - K \hat{h}_b^{1/2},
	&&\text{where}&
	K &=
	\begin{cases}
		\frac{\ppar*{8 H (H+1)}^{1/2}}{H^2 + H + 2} 	& \text{for } 0 < H \leq 1,	\\
		1 								    	& \text{for } 1 \leq H,
	\end{cases}
\end{align}
\end{subequations}
and $0 < K \leq 1$. The monotonicity of the relationship between $\hat{h}_b$ and $\hat{u}_b$ implies that selecting either $\hat{h}_b$ or $\hat{u}_b$ uniquely determines the other and, consequently, the flow state. From \cref{eqn:ot:app_dEdu_chain,eqn:ot:app_dhdu} we obtain
\begin{align}
	\pdv{\Delta \hat{E}_b}{\hat{u}_b} &= - \frac{\hat{h}_b^{1/2}}{F_b^{1/3}} ( 1 - F_b^{2/3} ) ( F_b^{2/3} + K F_b^{1/3} + 1 )
\end{align}
where the last term in the product is strictly positive. Finally, we conclude that for a subcritical flow state $0 < \hat{u}_b < \hat{h}_b^{1/2}$, we have $0 \leq F_b < 1 $, and therefore the derivative is negative.

To investigate other parametrisations of the flow state we expand $\pdv*{\Delta \hat{E}_b}{\hat{u}_b}$ using the chain rule, which means we require the derivative of our new parametrisation with respect to $\hat{u}_b$.

First we parametrise with $F_b$, where
\begin{subequations}
\begin{align}
	\pdv{F_b}{\hat{u}_b} = \frac{2 + K F_b}{2 \hat{h}_b^{1/2}} > 0
\end{align}
thus specifying $F_b$ uniquely determines the flow state, and
\begin{align}
	\pdv{\Delta \hat{E}_b}{F_b} &= - \frac{2 \hat{h}_b}{( 2 + K F_b ) F_b^{1/3}} ( 1 - F_b^{2/3} ) ( F_b^{2/3} + K F_b^{1/3} + 1 )
\end{align}
\end{subequations}
which is negative for a subcritical flow state. Next we parametrise with $\hat{\mu}_b \eqdef \hat{u}_b - \hat{h}_b^{1/2}$, for which
\begin{subequations}
\begin{align}
	\pdv{\hat{\mu}_b}{\hat{u}_b} &= 1 + \frac{K}{2} > 0
\end{align}
thus specifying $\hat{\mu}_b$ uniquely determines the flow state, and
\begin{align}
	\pdv{\Delta \hat{E}_b}{\hat{\mu}_b} &= - \frac{2 \hat{h}_b^{1/2}}{(2 + K) F_b^{1/3}} ( 1 - F_b^{2/3} ) ( F_b^{2/3} + K F_b^{1/3} + 1 ),
\end{align}
\end{subequations}
which is negative for a subcritical flow state. Next we parametrise using $\hat{s}$, which for a $\beta$-shock is the shock speed. By
\begin{subequations}
\begin{align}\label{eqn:ot:app_diff_shockspeed_by_vel}
	\pdv{\hat{s}}{\hat{u}_b} &= \hat{h}_b^{1/2} \frac{K (\hat{s} - \hat{\mu}_b) + (1-K)\hat{h}_b^{1/2}}{\hat{h}_b - \hat{h}_i} > 0,
\end{align}
we find that specifying $\hat{s}$ uniquely determines the flow state, and
\begin{align}
	\pdv{\Delta \hat{E}_b}{\hat{s}} &= - \frac{\hat{h}_b - \hat{h}_i}{K (\hat{s} - \hat{\mu}_b) + (1-K)\hat{h}_b^{1/2}} \frac{1}{F_b^{1/3}} ( 1 - F_b^{2/3} ) ( F_b^{2/3} + K F_b^{1/3} + 1 ),
\end{align}\end{subequations}
which is negative for a subcritical flow state. Finally we consider a parametrisation relevant to \cref{sec:overtop_app_constsurface}, namely the decrease in surface elevation from the flow at the edge of the domain to the flow over the barrier, given by
\begin{align}
	\Delta\hat{h} &\eqdef \hat{h}_b - 1 - (\hat{u}_b \hat{h}_b)^{2/3}.
\shortintertext{Using that}
	\pdv{\Delta\hat{h}}{\hat{u}_b} &= - \frac{\hat{h}_b^{1/2}}{3 F_b^{1/3}} \ppar*{ 3KF_b^{1/3} + 2(1-KF_b) }
\end{align}
is negative for subcritical flow we find that $\Delta\hat{h}$ uniquely determines subcritical outflow states, and 
\begin{equation} \label{eqn:ot:diff_energy_wrt_height_drop}
	\pdv{\Delta \hat{E}_b}{\Delta \hat{h}} = 3 \frac{ ( 1 - F_b^{2/3} ) ( F_b^{2/3} + K F_b^{1/3} + 1 ) }{ 3KF_b^{1/3} + 2(1-KF_b) }
\end{equation}
which is positive for a subcritical flow state.

%% file: Sections/Appendix_PerturbingShocks.tex
\section{Effect of perturbations on a shock}	\label{sec:overtop_PerturbShock}

We consider a shock satisfying \cref{eqn:ot:SW_shock_con_general}, and perturb the values and speed as
\begin{align}
	u^{\pm} &= u_{0}^{\pm} + \delta u_{1}^{\pm} + \order{\delta^2},	&
	h^{\pm} &= h_{0}^{\pm} + \delta h_{1}^{\pm} + \order{\delta^2},	&
	s &= s_{0} + \delta s_{1} + \order{\delta^2},
\end{align}
where $u_{0}^{\pm}$, $h_{0}^{\pm}$, and $s_{0}$ are assumed to satisfy the shock conditions exactly. Substitution of the perturbations and equating at $\order{\delta}$ yields the system
\begin{subequations}\begin{align}
	\shock*{(u_0-s_0)h_1 + (u_1-s_1)h_0} &= 0	\\
	\shock*{(u_0-s_0)^2 h_1 + 2(u_0-s_0)(u_1-s_1)h_0 + h_0h_1} &=0
\end{align}\end{subequations}
Defining $\tilde{q}_0 \eqdef (u_{0}^{-} - s_0) h_{0}^{-} = (u_{0}^{+} - s_0) h_{0}^{+}$, this system can be rearranged to give
\begin{subequations}\label{eqn:ot:perturbshock_uh}\begin{align}
	s_1 &= \frac{\shock*{\frac*{\tilde{q}_0 h_1}{h_0} + h_0 u_1}}{\shock*{h_0}},	\\
	0 &= \shock*{\ppar*{\frac{\tilde{q}_0^2}{h_0^2} + h_0} h_1 + 2 \tilde{q}_0 u_1}.
\end{align}\end{subequations}

In \cref{sec:overtop_collision_far} it is helpful to have the perturbed shock conditions in terms of the perturbations to the characteristic invariants
\begin{align}
	\alpha^{\pm} &\eqdef u^{\pm} + 2 \sqrt{h^{\pm}} = \alpha_{0}^{\pm} + \delta \alpha_{1}^{\pm} + \order{\delta^2},	&
	\beta^{\pm}  &\eqdef u^{\pm} - 2 \sqrt{h^{\pm}} = \beta_{0}^{\pm}  + \delta \beta_{1}^{\pm}  + \order{\delta^2}.
\end{align}
Substituting for our expansions of velocity and depth we obtain
\begin{align}
	\alpha_{1}^{\pm} &= u_{1}^{\pm} + \frac{h_{1}^{\pm}}{\sqrt{h_{0}^{\pm}}},	&
	\beta_{1}^{\pm}  &= u_{1}^{\pm} - \frac{h_{1}^{\pm}}{\sqrt{h_{0}^{\pm}}}
\intertext{thus}
	u_{1}^{\pm} &= \frac{\alpha_{1}^{\pm} + \beta_{1}^{\pm}}{2},	&
	h_{1}^{\pm} &= \sqrt{h_{0}^{\pm}}\frac{\alpha_{1}^{\pm} - \beta_{1}^{\pm}}{2}.	&
\end{align}
Therefore the expressions needed for the shock conditions are
\begin{subequations}\label{eqn:ot:perturbshock_ab}\begin{gather}
	\frac{\tilde{q}_0 h_{1}^{\pm}}{h_{0}^{\pm}} + h_{0}^{\pm} u_{1}^{\pm} 
	= \frac{\sqrt{h_{0}^{\pm}}}{2} \ppar*{ \ppar*{u_{0}^{\pm} + \sqrt{h_{0}^{\pm}} - s_0} \alpha_{1}^{\pm} - \ppar*{u_{0}^{\pm} - \sqrt{h_{0}^{\pm}} - s_0} \beta_{1}^{\pm} },	\\
	\ppar*{\frac{\tilde{q}_0^2}{(h_{0}^{\pm})^2} + h_{0}^{\pm}} h_{1}^{\pm} + 2 \tilde{q}_0 u_{1}^{\pm}
	= \frac{\sqrt{h_{0}^{\pm}}}{2} \ppar*{ \ppar*{u_{0}^{\pm} + \sqrt{h_{0}^{\pm}} - s_0}^2 \alpha_{1}^{\pm} - \ppar*{u_{0}^{\pm} - \sqrt{h_{0}^{\pm}} - s_0}^2 \beta_{1}^{\pm} }.
\end{gather}\end{subequations}